\documentclass[preprint]{aastex}
\begin{document}

\newcommand{\kms}{\ensuremath{\mathrm{km}\,\mathrm{s}^{-1}}}
\newcommand{\mJy}{\ensuremath{\mathrm{mJy}\,\mathrm{beam}^{-1}}}
\newcommand{\etal}{et al.}
\newcommand{\LCDM}{$\Lambda$CDM}
\newcommand{\ML}{\ensuremath{\Upsilon_{\star}}}
\newcommand{\MLfix}{\ensuremath{\Upsilon_{\star}^{fix}}}
\newcommand{\MLfree}{\ensuremath{\Upsilon_{\star}^{\rm free}}}
\newcommand{\MLsps}{\ensuremath{\Upsilon_{\star}^{3.6}}}
\newcommand{\MLspl}{\ensuremath{\Upsilon_{\star}^{4.5}}}
\newcommand{\MLk}{\ensuremath{\Upsilon_{\star}^{K}}}
\newcommand{\MLb}{\ensuremath{\Upsilon_{\star}^{B}}}
\newcommand{\MLmax}{\ensuremath{\Upsilon_{\star}^{\rm max}}}
\newcommand{\Lsun}{\ensuremath{{L}_{\odot}}}
\newcommand{\Msun}{\ensuremath{{M}_{\odot}}}
\newcommand{\mass}{\ensuremath{{\cal M}}}
\newcommand{\magsq}{\ensuremath{\mathrm{mag}\,\mathrm{arcsec}^{-2}}}
\newcommand{\Lsundens}{\ensuremath{L_{\odot}\,\mathrm{pc}^{-2}}}
\newcommand{\surfdens}{\ensuremath{M_{\odot}\,\mathrm{pc}^{-2}}}
\newcommand{\cubedens}{\ensuremath{M_{\odot}\,\mathrm{pc}^{-3}}}

\long\def\Ignore#1{\relax}
\long\def\Comment#1{{\footnotesize #1}}


\title{High-resolution dark matter density profiles of THINGS dwarf
  galaxies: Correcting for non-circular motions}

\author{Se-Heon Oh\altaffilmark{1}, 
W.J.G.\ de Blok\altaffilmark{2},
Fabian Walter\altaffilmark{3}, 
Elias Brinks\altaffilmark{4}, and
Robert C.\ Kennicutt, Jr.\altaffilmark{5}}

\email{seheon@mso.anu.edu.au}
\email{edeblok@ast.uct.ac.za}
\email{walter@mpia.de}
\email{E.Brinks@herts.ac.uk}
\email{robk@ast.cam.ac.uk}

\altaffiltext{1}{Research School of Astronomy \& Astrophysics,
The Australian National University, Mount Stromlo Observatory,
Cotter Road, Weston Creek, ACT 2611, Australia}
\altaffiltext{2}{Department of Astronomy, University of Cape Town, Private Bag X3, Rondebosch 7701, South Africa}
\altaffiltext{3}{Max-Planck-Institut f\"ur Astronomie, K\"onigstuhl 17, 69117 Heidelberg, Germany}
\altaffiltext{4}{Centre for Astrophysics Research, University of Hertfordshire, College Lane, Hatfield, AL10 9AB, United Kingdom}
\altaffiltext{5}{Institute of Astronomy, University of Cambridge, Madingley Road, Cambridge CB3 0HA, United Kingdom}


\begin{abstract}
  We present a new method to remove the impact of random and small-scale
  non-circular motions from H{\sc i} velocity fields in (dwarf)
  galaxies in order to better constrain the dark matter properties for
  these objects. This method extracts the circularly rotating velocity
  components from the H{\sc i} data cube and condenses them into a
  so-called bulk velocity field.  We derive high-resolution ($\sim
  0.2$ kpc) rotation curves of IC 2574 and NGC 2366 based on bulk
  velocity fields derived from The HI Nearby Galaxy Survey (THINGS)
  obtained at the VLA. We compare the bulk velocity field
  rotation curves with those derived from the traditional
  intensity-weighted mean velocity fields and find significant
  differences.  The bulk velocity field rotation curves are
  significantly less affected by non-circular motions and constrain
  the dark matter distribution in our galaxies, allowing us to address
  the discrepancy between the inferred and predicted dark matter
  distribution in galaxies (the ``cusp/core'' problem). {\it Spitzer}
  Infrared Nearby Galaxies Survey (SINGS) 3.6 $\mu$m data, which are
  largely unaffected by dust in these systems, as
  well as ancillary optical information, are used to separate the
  contribution of the baryons from the total matter content.  Using
  stellar population synthesis models, assuming various sets of
  metallicity and star formation histories, we compute stellar
  mass-to-light ratios for the 3.6 $\mu$m and 4.5 $\mu$m bands.  Using
  our predicted value for the 3.6 $\mu$m stellar mass-to-light ratio,
  we find that the observed dark matter distributions of IC 2574 and
  NGC 2366 are inconsistent with the cuspy dark matter halo predicted
  by $\Lambda$ Cold Dark Matter models, even after corrections for
  non-circular motions. This result also holds for other assumptions
  about the stellar mass-to-light ratio. The distribution of dark
  matter within our sample galaxies is best described by models with a
  kpc-sized constant-density core.
\end{abstract}


\section{Introduction}
Cosmological Cold Dark Matter (CDM) simulations have been very
successful in describing the observed large-scale structures
in the universe (Spergel \etal\ 2003; Primack 2003).  They have,
however, been less successful in describing the observed dark matter
density profiles of galaxies at small radii.  The most commonly used
models (Navarro, Frenk \& White 1996, hereafter NFW; see also Moore
\etal\ 1998) predict that the dark matter density profile increases
towards the center as a power law $\rho\sim R^{\alpha}$ with
$\alpha\sim -1$ to $-1.5$, giving rise to a ``cusp'' feature in the
centers of galaxies. However, most observations do not confirm this
NFW profile, instead preferring a sizeable central constant
density-core with  $\alpha \simeq -0.2\pm0.2$ (de Blok
\etal\ 2001; de Blok \& Bosma\ 2002).  The ``cusp/core'' problem is
one aspect of the small-scale crisis in $\Lambda$ Cold Dark Matter
(\LCDM) cosmology; the other two being the missing dwarf galaxies and
the angular momentum problem (Klypin \etal\ 1999; Moore \etal\ 1999;
Navarro \etal\ 1995).

Considerable research has been devoted to this problem, and the apparent
inability of standard \LCDM\, simulations to produce dark matter
density profiles that match the observed profiles is now well-known
(Flores \& Primack 1994; Moore 1994; de Blok \etal\ 2001; de Blok \&
Bosma 2002; Weldrake \etal\ 2003; Simon \etal\ 2003; Gentile
\etal\ 2004; see also Swaters \etal\ 2003).

It has been argued that several observational systematic effects, such
as beam smearing, pointing offsets, and non-circular motions, could
affect the measured inner slope of the density profile, and may
therefore ``hide'' the signatures of cusps in the central parts of
galaxies (van den Bosch \etal\ 2000; Swaters \etal\ 2003; Hayashi \&
Navarro 2006).  The early rotation curves used to study the dark
matter density profiles in Low Surface Brightness (LSB) galaxies were
mainly based on H{\sc i} data with large beam-sizes, and beam-smearing
could potentially have flattened the observed slopes of the rotation
curves. However, high-resolution H$\alpha$ rotation curves show
consistency with the H{\sc i} rotation curves (de Blok \etal\ 2001;
McGaugh \etal\ 2001; Kuzio de Narray \etal\ 2006).  From this, it
seems that beam smearing effects are not significant enough to erase
the signature of the cusp.

When using one-dimensional long-slit spectra, potential cusps may also
be ``hidden'' by telescope pointing offsets (i.e., slit offsets) with
respect to the galaxy centers. However, de Blok \etal\ (2003) show
that rotation curves observed by independent observers on different
telescopes agree within the given errors and that pointing offsets are
unlikely to exceed $\sim 0.3\arcsec$ (see also Marchesini
\etal\ 2002; de Blok \& Bosma 2002; de Blok 2004; Gentile
\etal\ 2007).  The insignificance of beam smearing and pointing
offsets is also confirmed by high-resolution 2D optical velocity
fields of LSB galaxies (Kuzio de Narray \etal\ 2006).

One of the fundamental assumptions of most rotation curve studies is
that the tracers that are typically used for galaxy dynamics, such as
H{\sc i} or H$\alpha$, travel on circular orbits. Therefore, any
significant non-circularity or random motions will affect the results
derived by such studies. Non-circular motions are not only thought to
be due to star formation processes, but also due to bars, spiral
density waves and non-circular halo potentials. For one dwarf LSB
galaxy (DDO 47), Gentile \etal\ (2005) have quantified the
non-circular motions and found them to be $\sim 2$ \kms, i.e., too
small by about an order of magnitude to explain the \LCDM\,
discrepancy. An analysis of non-circular motions in 19 THINGS
galaxies (Trachternach \etal\ 2008) also finds similar results:
especially in the low-luminosity galaxies, non-circular motions are
too small to explain the observed discrepancies.

A different problem is that in order to isolate the dynamical
contribution of the dark matter, the mass distribution of the baryons
(gas and stars) needs to be determined first. For the neutral gas this
can be directly derived from the integrated H{\sc i} map, but for the
stars this is less trivial. It requires knowledge of the stellar
mass-to-light ratio (hereafter \ML) which depends on
several factors that are not well constrained, such as the amount of
extinction from dust, the star formation history, and the stellar
Initial Mass Function. Moreover, these factors are interdependent,
making \ML\ one of the parameters in galaxy mass modeling with the
largest uncertainty. Therefore, in many studies, minimum and maximum
disk assumptions are often used (van Albada \& Sancisi 1986): the
minimum disk assumes that the observed rotation curve is entirely due
to dark matter and thus gives a hard upper limit to the dark matter
properties in galaxies. The maximum disk hypothesis maximizes the
rotation contribution of the stellar disk and thus provides an upper limit
on \ML\ or, equivalently, a lower limit on the contribution of dark matter.
However, these assumptions, though they provide useful limits, are
not able to determine the exact amount of dark matter in galaxies.

In this paper, we investigate the distribution of dark matter in two
nearby dwarf galaxies, IC 2574 and NGC 2366, using observations
obtained as part of The H{\sc i} Nearby Galaxy Survey (THINGS;
Walter \etal\ 2008).  Dwarf galaxies are  dark matter
dominated and the highly resolved THINGS observations provide an
opportunity to trace in detail the overall dynamics and constrain the
dark matter distribution.
The distorted velocity contours in IC 2574 and NGC 2366 
indicate the presence of non-circular motions 
(see Walter \etal\ 2008; de Blok et al.\ 2008; Trachternach et al.\ 2008).
A key element of our analysis
is a new approach based on a Gaussian fit algorithm to decompose
the H{\sc i} profiles.  This method reduces and quantifies the effects
of random non-circular motions or localized distortions due to, for
example, H{\sc i} shells and thus helps to extract the velocity
component representing the underlying ``undisturbed'' kinematics.  In
this paper, we only focus on the small-scale non-circular motions due
to, e.g., star formation events and which are not related to
large-scale features such as non-axisymmetric shapes of potentials
etc.\ (see Trachternach \etal\ 2008 for a more extensive discussion on
these large-scale non-circular motions). The adopted distances of 
IC 2574 and NGC 2366 are 4.0 kpc and 3.4 kpc, respectively (Walter \etal\ 2008). 

We also constrain the mass of the stellar component of IC 2574
by using {\it Spitzer} Infrared Array Camera (IRAC) 3.6 $\mu$m 
data from the {\it Spitzer} Infrared Nearby Galaxies Survey 
(SINGS; Kennicutt \etal\ 2003). NGC 2366 is not part of the SINGS
sample and for this galaxy we retrieved 3.6 $\mu$m data from the \emph{Spitzer} 
archives. We combine these data with population synthesis models 
derived using the ``GALAXEV'' package of Bruzual \& Charlot (2003).
We compare mass models derived using our
best 3.6 $\mu$m $\Upsilon_{\star}$ values with models derived under
different assumptions, such as maximum or minimum disk.

The organization of this paper is as follows: in Section 2, we give a
general description of data used in this paper. We derive the H{\sc i}
rotation curves of IC 2574 and NGC 2366 using a new Gaussian
decomposition method in Section 3. We present our determinations of
the masses of the stellar components of IC 2574 and NGC 2366 in Section
4.  The derived mass models of IC 2574 and NGC 2366 are presented in
Section 5 and fitted using a combination of halo models and
$\Upsilon_{\star}$ assumptions.  In Section 6, we measure the inner
slopes of the mass density profiles of IC 2574 and NGC 2366 and the main
results of this paper are summarized in Section 7.

\section{Data}
The H{\sc i} Nearby Galaxy Survey (THINGS) is one of the largest H{\sc
  i} survey programs undertaken with the NRAO\footnote{The National
  Radio Astronomy Observatory is a facility of the National Science
  Foundation operated under cooperative agreement by Associated
  Universities, Inc.}  Very Large Array (VLA) and comprises
observations of 34 nearby galaxies.  It has high spatial
($\sim$$6\,\arcsec$) and spectral ($\leq$ 5.2 \kms) resolutions (Walter
\etal\ 2008). In our analysis we use the natural-weighted cubes. In
order to preserve the noise characteristics of the data, no residual
scaling, primary beam correction or blanking was applied. The channel
separation in the two galaxies discussed here, $\delta V_{\rm channel}$, is 2.6 \kms.  To
constrain the contribution of the stellar component to the total
kinematics we use 3.6 $\mu$m data from the {\it Spitzer} Infrared Nearby Galaxies 
Survey (SINGS) (Kennicutt \etal\ 2003) for IC 2574 and
from the {\it Spitzer} archives for NGC 2366. These 
can be used as a proxy for the distribution of the stellar population. 
The resolution of $\sim 4\,\arcsec$ of the 3.6 $\mu$m
images is comparable to the THINGS resolution. In addition, we use
ancillary optical $B$, $V$, and $R$ images taken with the 2.1m
telescope at Kitt Peak National Observatory (KPNO) as part of the SINGS
survey.  We show the data for IC 2574 and NGC 2366 in
Fig. 1.

\section{Bulk velocity fields}
\subsection{The bulk-motion extraction method}
There are several ways to define a velocity field.  These include
first-moment maps, single Gaussian, multiple Gaussian, or hermite
polynomial fits, as well as peak velocity fields (see de Blok \etal\
2008 for a detailed discussion). Some, such as first-moment maps, Gaussian fits,
and peak velocity fields, work well for single, symmetrical H{\sc i}
profiles, but fail to properly take into account the presence of
multiple velocity components. In the upper-left panels of
Fig. 2 and 3
we show the intensity-weighted mean first-moment (hereafter IWM)
velocity fields of IC 2574 and NGC 2366 (the ``standard'' velocity
fields). From the distorted contours in these velocity fields, it is
clear that there are strong non-circular motions in both galaxies. It
is also clear that the IWM velocity fields are not optimal for
deriving a rotation curve and/or correcting for the non-circular
motions.  One can try and capture these non-circular motions by
fitting multiple Gaussian components to the profiles.  However, even though it
is possible to model a non-Gaussian profile using multiple Gaussian
components, the standard method often implemented in astronomical
software packages will frequently fail as it tries to fit these
components simultaneously.  This usually requires too many free
parameters (i.e., amplitude, dispersion, and central position for each
component used), quickly exceeding the typical number of data points
in an H{\sc i} profile. In addition, the presence of noise can result
in the fit becoming sensitive to the values of the initial
estimates. However, even if profiles are decomposed perfectly, we are
still left with the fundamental problem that a decomposition into
Gaussian components does not give much insight as to which component
is more representative of underlying circular rotation of the disk and
which one represents non-circular motion or additional velocity
components along the line of sight. In order to address this problem
and extract the undisturbed ``bulk motion'' (i.e., the velocity most
representative of the undisturbed rotation) from the H{\sc i} data
cube of a given galaxy, we devised an alternative Gaussian
decomposition method.  This method minimizes the effects of localized
non-circular motions (such as those caused by star formation
processes) and extracts the circularly rotating components from the
H{\sc i} data cube.  The method consists of the following steps.

\subsubsection{Step I: Estimate the initial, approximate rotation curves}

As an initial step, an approximate rotation curve is derived from a
major axis position-velocity diagram. For this we need to derive the
position angle (PA) of the major axis, but also the center coordinates
(XPOS, YPOS), inclination (INCL) and systemic velocity (VSYS).

Initial estimates for XPOS, YPOS, PA, and INCL can be derived from
ellipse fits to isophotes in the H{\sc i} surface density-map
($0^{th}$ moment map), optical, or infrared images.  This gives
approximate but reasonable initial values. In this paper, we used
ellipse fits to the 3.6 $\mu$m images of IC 2574 and NGC 2366 at a
level of 0.03 MJy sr$^{-1}$ to determine these initial values. We show
the ellipse fits to the 3.6 $\mu$m images for IC 2574 and NGC 2366 in
the middle panels of Fig. 1. An estimate for
VSYS is determined by averaging the velocity values found near the
center position. Using the obtained geometric parameters, we perform
single Gaussian fits to the profiles found along the major axis in the
H{\sc i} data cube, and extract the rotation velocity (VROT). If
multiple peaks with different velocities exist at a given spatial
position along the major axis, we select one of them as the initial
estimate for the bulk motion, as determined by the average velocity at
neighboring positions.  We emphasize that at this stage we are only
determining a first approximation to the rotation velocity and any
errors introduced by an incorrect selection of a velocity component
will be corrected in subsequent steps.

\subsubsection{Step II: Create the model velocity field for the bulk motion}

We now create an artificial velocity field using the previously
obtained geometrical parameters.  As we are at this stage only
interested in the overall shape of the rotation curve, we approximate
the values of VROT obtained above with a fifth-order polynomial.  The
model velocity field will be used as initial estimates for extracting
the bulk motions.  The model fields for IC 2574 and NGC 2366 are shown
in the top-right panels of Figs. 2 and 3, respectively.  Again, we stress that
these model velocity fields are just first approximations. They are
merely used to set the initial values for carrying out the multiple
Gaussian decomposition described in the following step.

\subsubsection{Step III: Perform the single Gaussian fit}

We now perform single Gaussian fits to extract the typical velocities
of all profiles in the data cube, taking into account the presence of
possible secondary components. For this, we developed a program
written in C. As is illustrated in the middle panel of
Fig. 4, if a profile consists of two Gaussian
components, the primary (i.e., strongest intensity) Gaussian component
is extracted as shown, taking into account the effect of the secondary
component on the fit. To do this, the program only uses data points
that are least affected by the secondary component (as shown by the
filled circles in the middle panel of Fig. 4).

We can judge the effect of the secondary component by making Gaussian
fits to both halves of the primary Gaussian component separately (as defined
with respect to the position of the peak flux). The fit to the side
containing the secondary component will result in a larger dispersion
compared to the side with only a single component.  This difference
can be used to indicate the approximate location of the secondary
component in a profile.

\subsubsection{Step IV: Extract the bulk velocity field via the
  multiple Gaussian decomposition}

In the last step, we assess whether the extracted velocities of the
profiles in $\it Step\,III$ are acceptable as representative of the
underlying circular rotation of the disk. For this, we compare those
velocities with the model velocity fields derived in $\it step\,II$.
If at some position the velocity difference ($\Delta V$) is less than
a prescribed limit (here we use $3\,\delta V_{\rm channel}$), we
select the velocity derived in $\it step\,III$ for the primary
component as the velocity of the bulk motion at that position.

However, if the difference is larger, this suggests that a better
velocity can be found by considering a secondary component, and we
proceed as follows.  The primary Gaussian component of the profile in
$\it step\,III$ is subtracted from the original profile and
subsequently a single Gaussian fit is made to the residual profile,
yielding a second Gaussian component. This is illustrated in the
right-most panel of Fig. 4. When performing
this second Gaussian fit, the model velocity field created in $\it
step\,II$ is used to determine an initial fit value.  We compare the
velocity derived for the secondary component with the model velocity
field derived. Likewise, we choose this velocity as the bulk velocity
at this position, if the velocity difference $\Delta V$ is less than a
certain limit (where we again use $3\,\delta V_{\rm channel}$).  In
cases where the velocity difference is larger, and no satisfactory
primary or secondary component is found, we simply put a blank value
in that position.  We also note that the extracted fits that satisfy
our $\Delta V < 3\,\delta V_{\rm channel}$ criterion, are only defined
as the ``bulk velocity'' component if they have a significant flux. We
demand a peak flux $> 3\sigma$, where $\sigma$ is the $\it rms$ noise
in the cube.  We thus substitute a blank value into the bulk velocity
field at positions where the extracted velocity shows strong
deviations or where no significant emission can be found.  This means
that our method does not ``invent'' data where there are none. Using
these procedures we have now extracted the first approximate bulk
velocity field from the H{\sc i} data cube.

We now repeat the entire process and go back to $\it
step\,I$. However, rather than using the major axis estimates, we now fit a
full ``tilted ring model'' (Begeman 1989) to the extracted bulk
velocity field obtained in $\it step\,IV$ and derive a second
approximation to the rotation curve parameters (XPOS, YPOS, VSYS, PA,
INCL, and VROT).  Using these newly determined tilted ring parameters,
we construct a second, improved model velocity field for the bulk
motion, which is less affected by additional components along the line
of sight, and then proceed to $\it steps\,II$, $III$, and $IV$.  In
this way, we iterate the above procedures until the mean difference
(after one round of $3\sigma$ outlier rejection) between successive
velocity fields is less than $3\,\delta V_{\rm channel}$.  We have
tested this method using different initial conditions (model velocity
fields) and find that the result is independent of these conditions.
Empirically it was determined that three or four iterations lead to a
stable result. The procedure is summarized in
Fig. 5.

The bulk velocity method works particularly well for galaxies 
in which non-circular motions form a significant part of the 
rotation velocity. For galaxies with high rotation velocities,
the method produces results very similar to the hermite 
polynomial method for constructing velocity fields described in 
de Blok et al.\ (2008).

\subsection{The bulk velocity fields of IC 2574 and NGC 2366}

We applied this method to IC 2574 and NGC 2366. The final extracted
bulk velocity fields are shown in the bottom-left panels of
Figs. 2 and 3, respectively. Most disturbances seen
in the IWM velocity fields of both galaxies are removed, especially in
the central regions of IC 2574 and the north-western part of NGC
2366. In general a warp significantly changes the position angle 
of the kinematical major axis as a function of radius and is found 
in the outer parts of a galaxy (Bosma 1978).
No significant changes of the derived position angles, as will be shown in Section 3.3, 
of the tilted ring models of IC 2574 and NGC 2366 are
shown in regions where most disturbances are seen. 
Therefore, these disturbances are not likely to be from a warp.
As a bonus, we also obtain the velocity fields of the strong
non-circular motions, which are shown in the bottom-right panels of
Figs. 2 and 3. These contain the velocities of the
primary components at the positions where these primary components
were found to track the non-circular motions, i.e., this is not a
residual velocity field. In order to visualize how well the extracted
bulk velocity field traces the rotation of the gas, we use azimuthal
position-velocity diagrams measured along ellipses defined using the
tilted ring models. In a quiescent galaxy with only circular motions,
one expects the observed velocities to follow a cosine as a function
of azimuth. Any regions containing non-circular motions deviate from
the cosine curves and we can therefore use the position-velocity
diagrams to assess whether the bulk-motion extraction method is able
to properly decompose the multiple profiles and identify the bulk
motion component properly (see Stil \& Israel 2002 for an analysis
which uses these azimuthal position-velocity diagrams to quantify
non-circular motions).

In Figs. 6 and 7, we
show the azimuthal position-velocity diagrams of IC 2574 and NGC 2366,
respectively. In Fig. 6, strong non-circular
motions in IC 2574 are easily identified at $\theta \sim 320^{\circ}$
at most radii. For example, in the region at $R \sim 228\arcsec$ and
$\theta \sim 320^{\circ}$, two clearly separated profiles are visible.
This is where the IWM velocities are strongly affected by non-circular
motions. These non-circular motions are caused by the effects of the
supergiant shell in IC 2574 (discussed in Walter \etal\ 1998).  In
contrast, the bulk velocities extracted by our method follow a
sinusoidal line tracing the underlying bulk motion of
IC 2574. Similarly, for NGC 2366 in Fig. 7 we
can see the presence of strong non-circular motions at $\theta \sim
270^{\circ}$ at most radii.  The IWM velocities, affected by
non-circular motions, display a non-symmetric shape in the
position-velocity diagram in regions with $R > 300\arcsec$ and
$\theta \sim 270^{\circ}$. The extracted bulk velocities again have
a sinusoidal-shape, indicating that the bulk velocity field for
NGC 2366 indeed traces the underlying rotation.

\subsection{H{\sc i} rotation curves}

We derived the rotation curves of IC 2574 and NGC 2366 using the bulk
velocity fields obtained in Section 3.2. In order to quantify the
difference with the IWM velocity field, we also derive the rotation
curves from the latter velocity fields.

We used the GIPSY task {\sc rotcur} to make tilted ring fits to the
bulk and IWM velocity fields.  The initial estimates of the ring
parameters, such as XPOS, YPOS, PA, and INCL were determined by an
ellipse fit to the IRAC 3.6 $\mu$m images for IC 2574 and NGC 2366
(except VSYS which was derived from the H{\sc i} global
profile). After fixing the position of the dynamical center (XPOS,
YPOS) and VSYS, we run {\sc rotcur} with PA and INCL as free
parameters. We made low-order polynomial fits (usually of fifth order)
to the PA and INCL distributions to describe the large-scale
variations with radius. We fixed PA and INCL to these fit values and
ran {\sc rotcur} again with XPOS, YPOS, and VSYS as free parameters.
From this, we radially average the XPOS, YPOS, and VSYS and
obtain more fine-tuned mean values. These values were fixed again
and {\sc rotcur} was run once more leaving the PA and INCL free.  We
iterated the {\sc rotcur} task in this way until all ring parameters
converged. For the last step, we fixed all ring parameters except VROT
and derived the final rotation curves of IC 2574 and NGC 2366.  The
derived rotation curves are shown in
Figs. 8 and 9 and the tilted ring parameters are
summarized in Table 1.

In the case of IC 2574 we found that the maximum difference between
rotation velocities as derived from the IWM and bulk velocity fields
is about $\sim$14 \kms\, at a radius of around 7 kpc, as shown in
Fig. 10 and Fig. 11 and this is a significant difference.
Note also the decreased scatter in the tilted ring
parameters, compared to the IWM tilted ring model presented in
Fig. 10. This can be considered as
circumstantial evidence that the bulk velocity field is less affected
by non-circular motions and therefore traces the underlying rotation
of IC 2574 more accurately than the IWM velocity field.  

We note the large difference between the bulk velocity field rotation
curve and the rotation curve derived by Martimbeau \etal\ (1994) in an
earlier study (Fig. 11). Martimbeau
\etal\ (1994) used an IWM velocity field and adopted a large value for
the inclination ($\sim$$75\,^{\circ}$), both of which are responsible
for this difference.

We compare the rotation curve of NGC 2366 derived from the bulk
velocity field with the one derived from the IWM velocity field in
Fig. 12.  The derived geometrical
tilted ring parameters from both bulk and IWM velocity fields are
consistent with each other, except for small differences in the
position angle.  We do find a significant velocity difference in the
outer parts.  Larger non-circular motions thus exist in the outer
parts of NGC 2366 confirming the impression given by the bottom-right
panel of Fig.3.

Fig. 13 compares our bulk rotation curve of
NGC 2366 with previous determinations by Hunter \etal\ (2001) and
Swaters (1999). The latter curve agrees with the current data out to
5.5 kpc. The Hunter \etal\ (2001) curve is systematically lower than
our bulk curve. A partial explanation is that they used a higher
inclination of $65^{\circ}$, but this difference is not large enough
to explain the entire discrepancy.  Though Hunter \etal\ (2001) do not
explicitly state how their velocity field was constructed, it is
likely that they used an IWM map, rather than one based on profile
fits. 

Declining rotation curves at the outer parts of a galaxy have
been often considered as an indication that the end of the dark matter distribution
is reached (Carignan \& Puche 1990; Casertano \& van Gorkom 1991;
Ryder \etal\ 1997; Carignan \& Purton 1998).
However, as shown in Fig. 13, the possibility
that a declining rotation curve at the outer parts of a galaxy 
is simply mimicked by non-circular motions cannot be ruled out.

The comparisons between IWM and bulk velocity results given above have
to some degree been qualitative. To really quantify the presence of
non-circular motions in both types of velocity fields and gauge the
effectiveness of the bulk velocity method in removing these effects,
we show here harmonic decompositions of the IWM and bulk velocity
fields of IC 2574. We refer to Trachternach et al.\ (2008) for an
extensive discussion of the method.
In summary, we used  the GIPSY task {\sc reswri} 
to decompose the velocity fields into 
sine and cosine terms and only kept 
the center position fixed during the harmonic expansion. We only included 
terms up to third order. The line-of-sight velocity then has the following form:
\begin{equation}
v_{los}(R)=v_{sys}+\sum_{n=1}^3\Biggl[c_{n}(R)\,\cos\, n\,\psi\, +\,s_{n}(R)\, \sin\, n\,\psi\Biggr] + {\rm resid.},
\end{equation}
where $c_0$ represents the systemic velocity, and $c_1$ the rotation
velocity, the $c_2,c_3,s_1,s_2$, and $s_3$ components quantify non-circular
motions.
For us, the total amplitudes of the non-circular motions are of most
interest.
The median absolute amplitudes $\langle A \rangle$ of each component
were calculated by taking the median of $A_n(R)$, where
\begin{equation}
A_n(R)=\sqrt{c_n(R)^2+s_n(R)^2},
\end{equation}
for $n>1$, and
\begin{equation}
A_1(R)=\sqrt{s_1(R)^2},
\end{equation}
for $ n=1$ ($c_1$ is the rotation velocity).
As can be seen in the top-left panel of Fig. 14, the radially
averaged, median amplitudes for the $\mit n=1$, and $\mit n=3$ components are small
for both velocity fields.
For the second order ($\mit n=2$), however, the harmonic decomposition of the
bulk velocity field results in a much smaller amplitude than that
of the IWM velocity field.
The effect of the bulk velocity method can be seen even more clearly in
the top-center panel of Fig. 14. Here, we plot
the absolute amplitude of the non-circular motions using the following equation:
\begin{equation}
\langle A(R) \rangle=\sqrt{s_1(R)^2 +  c_2(R)^2  +  s_2(R)^2  +  c_3(R)^2  +  s_3(R)^2}
\end{equation}
The harmonic components found in the IWM velocity field show
large amplitudes between $150\,\arcsec < R < 350\,\arcsec$, 
whereas the results from the bulk velocity field are
consistent with regular circular rotation.

\subsection{The mass model for the gas component}

The rotation curves obtained from the bulk velocity field reflect the
total (baryonic + dark matter) kinematics of galaxies and in order to
say anything about the dark matter, we thus need to quantify the 
contribution by the baryons.

To derive the contribution of the gas component to the total
kinematics the radial H{\sc i} surface density distribution is
required. This is obtained from the integrated H{\sc i} map (natural-weighted) (Walter
\etal\ 2008) using the derived tilted ring model parameters. The H{\sc
  i} radial surface density is then used to derive the corresponding
contribution to the rotation velocity after correcting the surface
density for the presence of helium and metals (i.e., after scaling by
a factor of 1.4; de Blok \etal\ 2008). We use the task {\sc rotmod}
implemented in GIPSY and assume an infinitely thin disk. The rotation
velocities for the gas component of IC 2574 and NGC 2366 are presented
in the bottom-left panels of Figs. 19 and 20, respectively.

\section{Stellar component}
\subsection{The 3.6 $\mu$m surface brightness profile}
After removing bright foreground stars in the vicinity of IC 2574 and NGC 2366 
manually, the 3.6 $\mu$m surface brightness profiles are determined using the 
{\sc ellint} task in GIPSY and the tilted ring fit parameters derived in Section 3.3. 
We used the standard IRAC calibration which provides flux values in 
units of MJy sr$^{-1}$. To convert this to surface brightness units of \magsq\, we use:
\begin{equation}
\mu_{3.6\mu\rm m} = -2.5\times\log_{10}\Biggl[\frac{S_{3.6\mu\rm m}\times2.35\times10^{-5}}{\rm ZP_{3.6\mu\rm m}}\Biggr], 
\end{equation}
where $S_{3.6\mu\rm m}$ is the flux value of the 3.6 $\mu$m band in units 
of MJy sr$^{-1}$. $\rm ZP_{3.6\mu\rm m}$ is the IRAC zero magnitude 
flux density in Jy and has 280.9 (Reach \etal\ 2005).
The final surface brightness profiles for IC 2574 and NGC 2366 in the  
3.6 $\mu$m band are shown in the top-left panel of Fig. 15 and
Fig. 16, respectively.
\subsection{Determining the 3.6 $\mu$m and 4.5 $\mu$m \ML\ values}

In general, the \ML\ values at optical wavelengths are affected by many
factors, including dust, age, metallicity, IMF, and recent star
formation. These give rise to large uncertainties in the inferred \ML\
values with corresponding uncertainties in dark matter halo
parameters. To circumvent this problem usually some assumptions for
the value of \ML, such as minimum disk or maximum disk are used when
performing disk-halo decompositions (van Albada \& Sancisi 1986; see
discussion in de Blok et al.\ 2008).  However, the derived dark matter
halo properties derived using these minimum/maximum disk assumptions
are, as discussed above, only upper/lower limits of the properties
on the dark matter halo.

Optical colors of the disk are also often used to put further
constraints on \ML. A relation between optical colors (e.g., $\it
B-R$, $\it B-V$) and \ML\ has been found in earlier work (e.g., Bell
\& de Jong 2001).  However these previous studies do not provide the
\ML\ values for the IRAC bands, so we cannot use them here. We have
therefore calculated the \ML\ values in the   3.6 $\mu$m and 4.5
$\mu$m bands.  For this calculation we constructed stellar population
synthesis models, with various sets of metallicity and star formation
histories using the ``GALAXEV'' package of Bruzual \& Charlot
(2003). Assuming an age of 12 Gyrs (as found for the Local Group in
Whiting 1999), we find a well-defined relation between \MLk\ and \ML\
in the   3.6 $\mu$m and 4.5 $\mu$m bands as shown in
Fig. 17.  These can be parameterized as follows,
\begin{equation}
\MLsps = B^{3.6}\times \MLk + A^{3.6},
\end{equation}
for the   $3.6\mu m$ band and,
\begin{equation}
\MLspl = B^{4.5}\times \MLk + A^{4.5},
\end{equation}
for the   $4.5$ $\mu$m band with  coefficients, 
$A^{3.6}$, $B^{3.6}$, $A^{4.5}$, and $B^{4.5}$ as given in Table 2.
Bell \& de Jong (2001; their Table 4) give the relations between 
the \MLk\ and optical colors and we have:
\begin{equation}
\log_{10}(\MLk) = b^{K}\times{\rm Optical\,Color} + a^{K},
\end{equation}
where $a^{K}$ and $b^{K}$ are also given in Table 2 for the appropriate colors.

Combining Eq. 6 with Eq. 8, adopting 20\% solar metallicity (Miller \&
Hodge 1996) and a scaled Salpeter IMF cutting off the stars less
massive than $\sim0.35$ \Msun\,(Bell \& de Jong 2001), we calculated
\ML\ for IC 2574 in the 3.6 $\mu$m band. As mentioned in Section
2, we use the ancillary optical $\it B$, $\it V$ and $\it R$ images to
derive the radial color distribution of IC 2574.  In the outer parts of
the galaxy, we extrapolate the surface brightness profile using
exponential fits. From these fits we extract optical colors ($\it B-R$
and $\it B-V$) of IC 2574 as shown in the top-right panel of
Fig. 15. The colors become bluer with increasing
radius, consistent with earlier results found for spiral galaxies (de
Jong 1996). In general, these color gradients can be explained by
different star formation histories and thus different present-day
stellar populations, with relatively older populations in the inner
parts than in the outer parts. This population change therefore also
implies radial changes in \ML\ in galaxies. In order to reflect this
\ML\ variation, we take the color gradient into account when
determining \ML. The resulting trends are shown in the bottom-left
panel of Fig. 15.  To determine our final \ML\
values, we  simply average the \MLk\ values as derived from $\it
B-R$ and $\it B-V$, respectively.  The average \MLsps\ used for the
final mass model of the stellar component of IC 2574 is shown in the
bottom-right panel of Fig. 15 (gray solid line).

Similarly, we calculated \ML\ values in the 3.6 $\mu$m band for NGC 2366 
using Eqs.\ 6 and 8. For NGC 2366, we used a 10\% solar metallicity 
(Hunter \etal\ 2006) and a scaled Salpeter IMF. We used the optical 
color ($\it B-V$) of NGC 2366 given in Hunter \etal\ (2006) and
found the coefficients of $a^{K}$ and $b^{K}$ as shown in Table 2.
We used a constant average color, ($\it B-V$)=0.31, since 
the radial $\it B-V$ color distribution given in Hunter \etal\ (2006)
is nearly constant (except where the super-giant H{\sc ii}\ region 
of NGC 2363 is located at a radius $\sim$1 kpc). 
From this we obtained \MLsps\,=0.33 for NGC 2366 as given in Table 2.
See also de Blok \etal\ (2008) for a comparison between the stellar 
disk masses of a number of THINGS galaxies derived using our method 
and the approach adopted by Leroy \etal\ (2008). 

The 3.6 $\mu$m emission may contain contributions from PAH
features but, in the case of the dwarfs, the contribution of PAHs is
not likely to be very high (see Walter \etal\ 2007 for a study of the
SINGS dwarfs). The 3.6 $\mu$m emission is likely to be a good proxy for the old
stellar population even if there can be contributions
from the intermediate-age population like AGB stars.
Although the 3.6 $\mu$m images thus provide a virtually
dust-free picture of the stellar component of a galaxy, the optical
colors that are implicitly used to determine \MLsps\ are possibly affected by
dust. However, Bell \& de Jong (2001) show that \MLk\ is only weakly
dependent on optical color, in contrast with \MLb.  This is
illustrated by the low values of $b^{K}$ in the IC
2574 \MLk\ relation ($\sim0.6$ for $\it B-V$ and $\sim0.4$ for $\it
B-R$), especially when compared with the much higher $b^{B}$ values listed in Bell
\& de Jong (2001).  The precise value of \MLsps\ is thus not very
sensitive to variations in $\it B-V$ or $\it B-R$.

\subsection{The mass model for the stellar components}

The 3.6 $\mu$m surface brightness profiles derived in Section 4.1 are 
in units of \magsq, and must, using the values of
$\Upsilon_{\star}$ derived in Section 4.2, still be converted to a mass
density profile in units of \surfdens.  The conversion to mass surface
density is calculated with the following formula:
\begin{equation}
\Sigma[\surfdens] = \MLsps \times 10^{-0.4\times(\mu_{3.6\mu\rm m} -– C^{3.6})},
\end{equation}
where $C^{3.6}$ is a constant for converting \magsq\, to \Lsundens.
We have:
\begin{eqnarray}
-2.5\log_{10}(1.0\,\Lsundens) &\simeq& -2.5\log_{10}(1.0\,\Lsun) + 21.56 \nonumber\\
&=& M_{\odot}^{3.6} + 21.56 \nonumber\\
&=& C^{3.6} 
\end{eqnarray}
where $M_{\odot}^{3.6}$ is the absolute solar magnitude 
in the   3.6 $\mu$m band. To calculate this, we perform the following steps.
\\
\\
\noindent \textbf{I. The apparent magnitude of the Sun}
\\
\\
\noindent The apparent magnitude of the Sun, $m_{\odot}^{\lambda}$, in the   3.6 $\mu$m or 4.5 $\mu$m band is given by:
\begin{equation}
m_{\odot}^{\lambda} = -2.5\log_{10}\Biggl[\int_{0}^{\infty}{\rm d}\lambda R(\lambda)f_{\odot}^{\lambda} \Biggr] + 2.5\log_{10}\Biggl[\int_{0}^{\infty}{\rm d}\lambda R(\lambda)f_{\alpha\,\rm Lyr}^{\lambda}\Biggr],
\end{equation}
where $R(\lambda)$ is the filter response function and 
$f_{\odot}^{\lambda}$ and $f_{\alpha\,\rm Lyr}^{\lambda}$ are the 
spectral energy distributions of the Sun and $\alpha$ Lyr respectively (Fukugita \etal\ 1995). 
The Kurucz model (1992) for $\alpha$ Lyr has $T_{\rm eff}$=9400K, 
$\log_{10}g$=3.90, and $\log_{10}Z=-0.50$. The convolved spectral 
energy distributions of the Sun and $\alpha$ Lyr using the IRAC filter 
response functions in the 3.6 $\mu$m and 4.5 $\mu$m bands are shown 
in Fig. 18. We compute the areas below these convolved 
spectral energy distributions of the Sun and $\alpha$ Lyr.
Using Eq. 11, we then obtain the apparent magnitude of the Sun 
as $-28.33$ and $-28.30$ in the   3.6 $\mu$m and 4.5 $\mu$m bands, respectively.
\\
\\
\noindent \textbf{II. The absolute magnitude of the Sun in the IRAC bands}
\\
\\
\noindent We obtain the absolute magnitude of the Sun as follows,
\begin{equation}
m_{\odot}^{\lambda}-M_{\odot}^{\lambda} = 5\log_{10}D_{\odot}-5,\nonumber
\end{equation}
where $D_{\odot} = 1/206625$ pc (distance to the Sun). From this, we find
\begin{equation}
M_{\odot}^{3.6} = m_{\odot}^{3.6} + 31.57 = 3.24\nonumber
\end{equation}
\begin{equation}
M_{\odot}^{4.5} = m_{\odot}^{4.5} + 31.57 = 3.27\nonumber
\end{equation}
From Eq. 10, $C^{3.6} = 24.80$, and using Eq. 9 we then convert 
the 3.6 $\mu$m surface brightness to a mass density in units of 
\surfdens. The final mass models for IC 2574 and NGC 2366 using 
the mass density for the stellar component are constructed with 
the {\sc rotmod} task in GIPSY. These are illustrated in 
Fig. 19 and Fig. 20.
For the stellar disk, we assume a vertical $sech^{2}$ scale-height 
distribution with $h/z_{0}=5$ (van der Kruit \& Searle 1981) where $h$ 
is the radial scale-length and $z_{0}$ is the scale-height of disk in kpc.
This $h/z_{0}=5$ ratio as determined in van der Kruit \& Searle (1981)
is based on a small number of galaxies but has, for a much larger sample,
been confirmed by Kregel \etal\ (2002), who find $h/z_{0}=4.8\pm1.3$.
From exponential disk fits to the 3.6 $\mu$m surface brightness distributions
(top-left panel of Fig. 15 and Fig. 16)
we obtain scale-height values of IC 2574 and NGC 2366 of $z_{0}=0.57$ and 
0.34 kpc, respectively.

\section{Dark matter distribution}
\subsection{Dark matter halo models}
The properties of the dark matter halo of a galaxy are usually quantified 
by using dark matter halo models. The residuals obtained by subtracting 
from the derived rotation curves those contributions corresponding to 
the stellar and gas components, are assumed to be due to the dark matter halo.
In this paper, we explore two models: the \LCDM\,NFW cusp-dominated halo 
and the pseudo-isothermal core-dominated halo. Their properties are given below.

\subsubsection{NFW dark matter halo model}
Navarro, Frenk \& White (1996, 1997; NFW) give a prescription 
for the dark matter distribution found in numerical simulations, 
based on the CDM\,paradigm. This so-called ``universal density 
profile'' has a cusp feature towards the galaxy center.
The profile has the form
\begin{equation}
\rho_{\rm NFW}(R) = \frac{\rho_{i}}{(R/R_{s})(1+R/R_{s})^{2}},
\end{equation}
where $\rho_{i}$ is the initial density of the universe at the 
time of collapse of the halo and $R_{s}$ is the characteristic 
radius of the dark matter halo. The rotation velocity corresponding 
to the NFW halo density is given as
\begin{equation}
V_{\rm NFW}(R) = V_{200}\sqrt{\frac{\rm ln(1+cx)-cx/(1+cx)}{x[\rm ln(1+c)-c/(1+c)]}},
\end{equation}
where $c$ is the concentration parameter and defined as $R_{200}/R_{s}$.
$V_{200}$ is the rotation velocity at radius $R_{200}$ where the 
density contrast exceeds 200 and $x$ is defined as $R/R_{200}$.
This universal density profile can be approximated using 
two power laws: $\rho \propto R^{-1}$ at small radii and 
$\rho \propto R^{-3}$ at large radii. See de Blok \etal\ (2008) 
for a more extensive description.

\subsubsection{Pseudo isothermal dark matter halo model}
The spherical pseudo-isothermal halo model which is used in most 
of the early rotation curve studies is observationally motivated
and has a core-like constant density profile which can be described
as $\rho \propto R^{0}$ towards the galaxy center and $R^{-2}$ for large $R$. 
It has the following form,
\begin{equation}
\rho_{\rm ISO}(R) = \frac{\rho_{0}}{1+(R/R_{c})^{2}},
\end{equation}
where $\rho_{0}$ and $R_{c}$ are the core-density and core-radius
of the dark matter halo, respectively. This density profile gives the rotation velocity,
\begin{equation}
V_{\rm ISO}(R) = \sqrt{4\pi G\rho_{0}R_{c}^{2}\Biggl[1-\frac{R_{c}}{R}{\rm atan}\Biggl(\frac{R}{R_{c}}\Biggr)\Biggr]}.
\end{equation}
This approaches the asymptotic velocity at large radii given by
\begin{equation}
V_{\infty} = \sqrt{4\pi G\rho_{0}R_{c}^{2}}.
\end{equation}
\subsection{Mass modeling results}

We now construct mass models using the rotation curves derived from
the bulk velocity fields of IC 2574 and NGC 2366 taking into account
the distribution of baryons, and use these to fit NFW and
pseudo-isothermal halo models.  Given the proximity of IC 2574 and NGC
2366, the THINGS (natural-weighted) resolution of $\sim 12\arcsec$
corresponds to $\sim$ 200 pc, comparable to that of the simulations of
dwarf dark matter halos presented in Navarro \etal\ (2004). As
discussed earlier, differences between NFW and pseudo-isothermal dark
matter models are the most distinct in the inner parts of
galaxies. Also, as the tilted rings in the outer parts of our galaxies
are only partly filled with emission, we estimate the effects of these
outer asymmetric features by presenting fits to both the entire radial
ranges of the rotation curves, as well as fits to the inner parts only
(i.e., $R \le$ 7.5 kpc for IC 2574 and $R \le$ 6 kpc for NGC 2366).
 
When performing the fits we also explore classical \ML\ assumptions,
such as maximum disk, minimum disk, minimum disk + gas (the stellar
component is ignored, but the contribution of the gas is taken into
account), in addition to the model \MLsps\ values we calculated. We
also attempted a fit with \MLsps\ as a free parameter. The results are
given in Figs. 21 and 22, and Tables 3 and 4.  The
best-fitting unconstrained values for \ML\ differ somewhat from
those predicted by the population synthesis modeling described earlier
(except for the IC 2574 pseudo-isothermal fit). The negative values of
\ML\ that the curves seem to prefer are obviously unphysical.  An
extensive discussion of this behavior is given in de Blok et al.\
(2008).

We find that the pseudo-isothermal halo provides a better fit than the
NFW halo. NFW models resulted in inferior fits, independent of which
value for \ML\ was adopted. The preferred $c$ values (concentration
parameter) are all negative with extremely large uncertainties. When
making the final fits, we fixed these values to be $c=0.1$. See de
Blok \etal\ (2008) for a discussion on the $c$-values found within the
THINGS sample.  In contrast, the pseudo-isothermal halo models provide
reasonable halo parameters, except for the IC 2574 maximum disk
case. However, we found that the fit is very sensitive to the choice
of \MLmax\ for the maximum disk and sensible parameters can be
obtained by slightly lowering \MLmax. 
  These results show that the NFW distribution is not an
appropriate fitting function for our galaxies and that the implied
dark matter distributions of the two dwarf galaxies, IC 2574 and NGC
2366, instead show a sizeable central constant-density core, which can
be well approximated by the pseudo-isothermal halo model.

We now compare the H{\sc i} rotation curves and mass modeling based on 
the bulk velocity fields with a similar analysis done using the 
IWM velocity fields to examine the effect of non-circular motions 
on the dynamics of galaxies.
First, for convenience, we introduce some notation.
We use $V_{\rm bulk}$ to indicate the rotation curve derived from
tilted ring fits using the bulk velocity field, as described 
in Section 3.3.
We use $V^{\prime}_{\rm IWM}$ to describe the rotation curve derived assuming
the geometrical parameters from the bulk velocity field tilted ring model
but applied to the IWM velocity field.
Finally, we use $V_{\rm IWM}$ to describe the rotation curve derived
using a tilted ring model as derived from and applied to the IWM velocity field.

For a galaxy affected by non-circular motions, the traditional
IWM velocity field will be distorted. In general, non-circular 
motions as present in galaxies, are likely to make the observed 
rotation velocity fall below the circular velocity (Rhee \etal\ 2004).
This is indeed happening in our galaxies as indicated in 
Fig. 23, where $V_{\rm IWM}$ is 
significantly lower than $V_{\rm bulk}$. The velocity difference 
between $V_{\rm bulk}$ and $V^{\prime}_{\rm IWM}$ can only be 
attributed to differences in the bulk and IWM velocity fields,
as we use otherwise identical geometrical parameters.

In the case of IC 2574, a distinct velocity difference
is seen in the central region where the IWM velocity field is 
distorted by non-circular motions (Fig. 23, left column). Such a velocity difference is also found in NGC 2366, 
particularly in the outer parts (see the Fig. 23, right column).

In practice, this underestimate of the rotation velocity results in a
decreased dynamical contribution of the dark matter to the total
dynamics, as the contribution of stellar and gas components are fixed
for any given $\Upsilon_{\star}$ assumption.  Consequently,
non-circular motions decrease the role of dark matter with respect to
the visible matter.  In the bottom-left panel of
Fig. 23 the rotation velocity due to the
stellar component using the model \MLsps\ values for IC 2574 already
exceeds $V^{\prime}_{\rm IWM}$ and $V_{\rm IWM}$ in the inner
parts even without the dynamical contribution of the gas
component. The $V_{\rm bulk}$ rotation curve, however, can in the
inner parts not only accommodate the rotation of the stellar component, but
also that of the gas component.

\section{Dark matter mass density profiles}
An intuitive way to illustrate the dark matter distribution 
of a galaxy is to calculate the mass density that corresponds 
to the observed rotation velocity. The observed rotation velocity 
is converted to mass density by assuming a spherical 
mass distribution (i.e., $\nabla^{2}\Phi$ = $4\pi G\rho$, $\Phi = -GM/R$).
For the inversion, we use the following formula (de Blok \etal\ 2001),
\begin{equation}
4\pi G\rho(R) = 2\frac{V}{R}\frac{\partial V}{\partial R} + \Biggl(\frac{V}{R}\Biggr)^{2},
\end{equation}
where $V$ is the rotation velocity observed at radius $R$.  This
direct conversion is only valid under the assumption that the observed
rotation velocity is entirely due to the dark matter component, as is the
case with a minimum disk (i.e., maximum halo).  In general the minimum
disk assumption is a good description for dwarf galaxies and LSB
galaxies (de Blok \etal\ 2002), and we therefore also apply this
assumption to IC 2574 and NGC 2366.  While one could, in principle,
make an explicit correction for the rotational contribution of the
baryons, for the crucial innermost parts this involves using the
difference between two small numbers, which can lead to wildly
fluctuating values of the derivative in Eq. 20. The minimum disk
assumption thus yields a safe and robust upper limit on the dark
matter properties and in particular to the steepness of the inner
slope (de Blok \& McGaugh 1997; de Blok \etal\ 2001).

We follow the method described in de Blok \etal\ (2002) to determine
the slope of the inner component of the mass density profile.  We
measure the slopes of the inner parts ($R < 1.2$ kpc) of IC 2574 and
NGC 2366 using a least squares fit and find the values of the slopes
to be $\alpha = +0.13\pm 0.07$ for IC 2574 and $\alpha = -0.32\pm 0.10$ 
for NGC 2366, respectively. These are shown in
Fig. 24 and are in good
agreement with the earlier result of $\alpha= -0.2\pm 0.2$ (de Blok
\etal\ 2001, de Blok \& Bosma 2002) for a larger sample of LSB
galaxies.
These flat slopes
thus imply that the dark matter distributions of IC 2574 and NGC 2366
are well characterized by a sizeable constant-density core.  The mass
density profiles for the best-fit minimum disk NFW and pseudo-isothermal
models are also over-plotted in Fig. 24. 
The pseudo-isothermal halo follows the observed mass density profile 
most closely. This is in sharp contrast with the steep slope predicted 
by the NFW profile.

We plot the value of the inner-slope $\alpha$ of the mass-density 
profile against the observed radius of the innermost point $R_{\rm inner}$
in Fig. 25. The result is consistent with the 
earlier results by de Blok \etal\ (2002). In conclusion, the observed dark 
matter distributions of IC 2574 and NGC 2366 are both best described by the 
pseudo-isothermal halo model with a constant-density core.

\section{Summary}
We have presented mass models for the nearby dwarf galaxies, IC 2574
and NGC 2366, derived using the high-resolution data from THINGS.
These high-resolution data do not suffer from beam smearing, 
have a well-defined dynamical center and enable us to examine in detail the dark matter
distribution of these galaxies. To minimize the effects of random
non-circular motions on the derived kinematics of a galaxy, we developed
a new Gaussian decomposition method and used this to construct a
``bulk'' velocity field of IC 2574 and NGC 2366, showing the
underlying undisturbed rotation.

The random, non-circular motions of IC 2574 and NGC 2366, visible as
distortions in the velocity contours of the traditional IWM velocity
fields, were largely removed in the newly constructed bulk velocity
fields. Comparing the H{\sc i} rotation curves derived from the bulk
and IWM velocity fields, we find that the rotation velocities derived
from the IWM velocity fields are significantly lower than those from
the bulk velocity fields. In addition, non-circular motions of NGC 2366 are
likely to be responsible for the declining rotation velocities derived from the
IWM velocity fields in the outer parts.

Combining optical and SINGS 3.6 $\mu$m data,
we quantify the dynamical contribution of the stellar component to the
total kinematics. For this we compute \ML\ values based on the
Bruzual and Charlot (2003) population synthesis models for the 3.6
$\mu$m and 4.5 $\mu$m bands.

We have fitted NFW and pseudo-isothermal dark matter halos to the derived
rotation curves, taking into account the contributions due to stars
and gas. We found that the pseudo-isothermal halo provides a better
fit to the observations than the NFW halo.  We use the derived mass
density profile to determine the value of the inner slope. The
measured slopes are $\alpha = +0.13\pm 0.07$ for IC 2574 and
$\alpha = -0.32\pm 0.10$ for NGC 2366, compared to the NFW
model which predicts $\alpha \sim-1$. 

The dark matter distributions of
IC 2574 and NGC 2366 are well described by the
pseudo-isothermal model ($\alpha \sim0$) with a sizeable central
constant-density core. These results are not affected by systematic
effects due to lack of resolution or pointing offsets, take into account the
effects of non-circular motions and use a well-constrained model for
\ML.

\acknowledgements
The work of WJGdB is based upon research supported by the South  
African Research Chairs Initiative of the Department of Science and  
Technology and National Research Foundation.
EB gratefully acknowledges financial support through an EU Marie Curie
International Reintegration Grant (Contract No. MIRG-CT-6-2005-013556).
This research has made use of the NASA/IPAC
Extragalactic Database (NED) which is operated by the Jet Propulsion
Laboratory, California Institute of Technology, under contract with the
National Aeronautics and Space Administration.
This publication makes use of data products from the Two Micron All
Sky Survey, which is a joint project of the University of
Massachusetts and the Infrared Processing and Analysis
Center/California Institute of Technology, funded by the National
Aeronautics and Space Administration and the National Science
Foundation.


\begin{deluxetable}{lcccccc}
\tablewidth{0pt}
\tablecaption{Parameters of tilted ring models of IC 2574 and NGC 2366}
\label{TILTED_RINGS}
\tablehead{
\colhead{Name}   &  \colhead{$\alpha$ (2000.0)}    &  \colhead{$\delta$ (2000.0)} &  \colhead{$V_{\rm sys}$} &  \colhead{$\langle i \rangle$} & \colhead{$\langle$PA$\rangle$} &  \colhead{$D$} \\ 
\colhead{}       & \colhead{(hh:mm:ss.s)} & \colhead{(dd:mm:ss.s)} & \colhead{(\kms)} & \colhead{($^{\circ}$)} & \colhead{($^{\circ}$)} & \colhead{(Mpc)} \\ 
\colhead{}& \colhead{(1)} & \colhead{(2)}   & \colhead{(3)}      & \colhead{(4)}    & \colhead{(5)}    & \colhead{(6)}}
\startdata
NGC 2366		& 07:28:53.4 & +69:12:51.1 & 104.0 & 63.8 & 39.8 & 3.4 \\
IC 2574		& 10:28:27.7 & +68:24:59.4 & 53.1  & 53.4 & 55.7 & 4.0 \\
\enddata
\tablecomments{
{\bf (1)(2):} Center positions.
{\bf (3):} Systemic velocity derived from a tilted ring fit using the bulk velocity field as described in Section 3.3.
{\bf (4):} Average inclination of the tilted ring model.
{\bf (5):} Average position angle of the tilted ring model.
{\bf (6):} Distances as given in Walter \etal\ (2008).}
\end{deluxetable}
\clearpage

\begin{deluxetable}{cccccccc}
\tablewidth{0pt}
\tablecaption{Coefficients for the \ML\ relations Eqs.~6--8}
\label{ML_PARAM}
\tablehead{
\colhead{$\lambda$} & \multicolumn{3}{c}{$A^{\lambda}$} && \multicolumn{3}{c}{$B^{\lambda}$}\\
&   & \colhead{(1)} &  &   &  & \colhead{(2)} &  
}
\startdata
3.6 $\mu$m  &  & $-0.05$   & & & & 0.92 & \\
4.5 $\mu$m  &  & $-0.08$   & & & & 0.91 & \\
\tableline
\tableline
\noalign{\vskip 2pt}
& \multicolumn{3}{c}{IC 2574} && \multicolumn{3}{c}{NGC 2366}\\
\cline{2-4} \cline{6-8}
\noalign{\vskip 2pt} 
Color   &$a^{K}$& $b^{K}$ & $\langle\MLsps\rangle$ & &$a^{K}$& $b^{K}$ & $\langle\MLsps\rangle$\\
& (3) & (4)  & (5)  &   & (6)  & (7) & (8) \\
\tableline
$B-V$    &$-0.59$& 0.60 & 0.44 & &$-0.60$& 0.72 & 0.33 \\
$B-R$    &$-0.67$& 0.42 &      & &  \nodata     & \nodata &      \\
\enddata
\tablecomments{
{\bf (1)(2):} Coefficients for the relations between \MLk\ and \ML\ in the   3.6 $\mu$m and 4.5 $\mu$m bands
derived in Section 4.2.
{\bf (3)(4):} Coefficients for the relations between \MLk\ and optical colors
given in Bell \& de Jong (2001), adopting 20\% solar metallicity and a scaled Salpeter IMF for IC 2574.
{\bf (5)(8):} Average \ML\ in the 3.6 $\mu$m band (\Msun/\Lsun).
{\bf (6)(7):} Coefficients for the relation between \MLk\ and optical color
given in Bell \& de Jong (2001), adopting 10\% solar metallicity and a scaled Salpeter IMF for NGC 2366.}
\end{deluxetable}
\clearpage

\begin{table*}
\scriptsize
\caption{Parameters of dark halo models for IC 2574}
\label{NFWHALO_IC2574}
\begin{center}
\begin{tabular}{@{}lrclccccrc}
\hline
\hline
\noalign{\vskip 2pt}
& \multicolumn{4}{c}{NFW halo (entire region)} && \multicolumn{4}{c}{NFW halo ($<$ 7.5 kpc)}\\
\cline{2-5} \cline{7-10}\\
\multicolumn{1}{c}{\ML\ assumption}   &  \multicolumn{1}{c}{$\langle\MLsps\rangle$}    &  \multicolumn{1}{c}{$c$}   &  \multicolumn{1}{c}{$V_{200}$}  &  \multicolumn{1}{c}{$\chi_{red.}^2$}    &&  \multicolumn{1}{c}{$\langle\MLsps\rangle$} &  \multicolumn{1}{c}{$c$}   &  \multicolumn{1}{c}{$V_{200}$}  &  \multicolumn{1}{c}{$\chi_{red.}^2$}\\
\multicolumn{1}{c}{(1)} & \multicolumn{1}{c}{(2)}   & \multicolumn{1}{c}{(3)}      & \multicolumn{1}{c}{(4)}    & \multicolumn{1}{c}{(5)}    && \multicolumn{1}{c}{(6)} & \multicolumn{1}{c}{(7)} & \multicolumn{1}{c}{(8)}  & \multicolumn{1}{c}{(9)} \\
\hline
Min. disk		& 0.00	& $<0.1$  &  674.6 $\pm$ 18.3   & 2.88   && 0.00    & $<0.1$   & 1213.6 $\pm$ ...  & 3.39\\
Min. disk+gas		& 0.00 	& $<0.1$  &   524.3 $\pm$ 51.7   & 1.65   && 0.00    & $<0.1$   & 1005.5 $\pm$ ... & 2.32\\
Max. disk		& 0.93	& $<0.1$  &   634.4 $\pm$ ...    & 2.33   && 0.93    & $<0.1$   & 353.8 $\pm$ ... & 1.63\\
Model $\Upsilon_{*}^{3.6}$ disk & 0.44  & $<0.1$  & 873.9 $\pm$ ... & 1.81&& 0.44    & $<0.1$   & 700.5 $\pm$ ... & 1.96\\
\MLfree\,disk		& $-0.03$ & $<0.1$ & 1107.7 $\pm$ ... & 1.66        &&         &          &                  &     \\
\hline\\
\noalign{\vskip 2pt}
 & \multicolumn{4}{c}{Pseudo-isothermal halo (entire region)} && \multicolumn{4}{c}{Pseudo-isothermal halo ($<$ 7.5 kpc)}\\
\cline{2-5} \cline{7-10} \\
\multicolumn{1}{c}{\ML\ assumption}    & \multicolumn{1}{c}{$\langle\MLsps\rangle$}   &  \multicolumn{1}{c}{$R_{C}$} &   \multicolumn{1}{c}{$\rho_0$}    & \multicolumn{1}{c}{$\chi_{red.}^2$}     && \multicolumn{1}{c}{$\langle\MLsps\rangle$} & \multicolumn{1}{c}{$R_{C}$}      & \multicolumn{1}{c}{$\rho_{0}$}  & \multicolumn{1}{c}{$\chi_{red.}^2$} \\
\multicolumn{1}{c}{(10)} & \multicolumn{1}{c}{(11)}   & \multicolumn{1}{c}{(12)}      & \multicolumn{1}{c}{(13)}    & \multicolumn{1}{c}{(14)}    && \multicolumn{1}{c}{(15)} & \multicolumn{1}{c}{(16)} & \multicolumn{1}{c}{(17)}  & \multicolumn{1}{c}{(18)} \\
\hline
Min. disk		& 0.00	&  5.77 $\pm$ 0.16       & 7.8 $\pm$ 0.2    & 0.25   && 0.00   & 5.69 $\pm$ 0.35       & 7.8 $\pm$ 0.3     & 0.26\\
Min. disk+gas		& 0.00 	&  4.61 $\pm$ 0.12       & 7.6 $\pm$ 0.2   & 0.16   && 0.00   & 3.88 $\pm$ 0.16       & 8.7 $\pm$ 0.3     & 0.13\\
Max. disk		& 0.93	&  27.52 $\pm$ 10.22       & 1.8 $\pm$ 0.1    & 0.30   && 0.93   & $...$               & 1.7 $\pm$ 0.9     & 0.39\\
Model $\Upsilon_{*}^{3.6}$ disk & 0.44  &  7.23 $\pm$ 0.30       & 4.1 $\pm$ 0.1    & 0.17   && 0.44   & 5.87 $\pm$ 0.55       & 4.5 $\pm$ 0.3     & 0.19\\
\MLfree\,disk		& 0.10  &  4.99 $\pm$ 0.34       & 6.7 $\pm$ 0.7    & 0.16          &&                                  &          &     \\
\hline
\end{tabular}
\medskip\noindent
\begin{minipage}{156mm}
\noindent
\\
{\bf Note.$\--$}
{\bf (1)(10):} The stellar mass-to-light ratio \ML\ assumptions. ``Model \MLsps\ disk'' uses the values derived from the population synthesis models in Section 4.2.
\MLfree\ has \MLsps\ as a free parameter.
{\bf (2)(6)(11)(15):} Average \ML\ in the 3.6 $\mu$m band (\Msun/\Lsun).
{\bf (3)(7):} Concentration parameter c of NFW halo model (NFW 1996, 1997).
{\bf (4)(8):} The rotation velocity (\kms)\,at radius $R_{200}$ where the density constrast exceeds 200 (Navarro \etal\ 1996).
{\bf (5)(9)(14)(18):} Reduced $\chi^{2}$ value.
{\bf (12)(16):} Fitted core-radius of pseudo-isothermal halo model (kpc).
{\bf (13)(17):} Fitted core-density of pseudo-isothermal halo model ($10^{-3}$ \cubedens).
{\bf ($...$):} blank due to unphysically large value or not well-constrained uncertainties.
\end{minipage}
\end{center}
\end{table*}
\clearpage

\begin{table*}
\scriptsize
\caption{Parameters of dark halo models for NGC 2366}
\label{NFWHALO_NGC2366}
\begin{center}
\begin{tabular}{@{}lrclccccrc}
\hline
\hline
\noalign{\vskip 2pt}
& \multicolumn{4}{c}{NFW halo (entire region)} && \multicolumn{4}{c}{NFW halo ($<$ 6.0 kpc)}\\
\cline{2-5} \cline{7-10}\\
\multicolumn{1}{c}{\ML\ assumption}   &  \multicolumn{1}{c}{$\langle\MLsps\rangle$}     &  \multicolumn{1}{c}{$c$}   &  \multicolumn{1}{c}{$V_{200}$}  &  \multicolumn{1}{c}{$\chi_{red.}^2$}    &&  \multicolumn{1}{c}{$\langle\MLsps\rangle$} &  \multicolumn{1}{c}{$c$}   &  \multicolumn{1}{c}{$V_{200}$}   &  \multicolumn{1}{c}{$\chi_{red.}^2$}\\
\multicolumn{1}{c}{(1)} & \multicolumn{1}{c}{(2)}   & \multicolumn{1}{c}{(3)}      & \multicolumn{1}{c}{(4)}    & \multicolumn{1}{c}{(5)}   && \multicolumn{1}{c}{(6)}  & \multicolumn{1}{c}{(7)} & \multicolumn{1}{c}{(8)}  & \multicolumn{1}{c}{(9)} \\
\hline
Min. disk        & 0.00	 & $<0.1$      & 901.5 $\pm$ 478.4   & 1.72	   &&  0.00	   &  $<0.1$       & 1600.5 $\pm$ $...$    & 2.35\\
Min. disk+gas    & 0.00	 & $<0.1$      & 727.8 $\pm$ $...$   & 1.08	   &&  0.00 	   &  $<0.1$        & 1136.6 $\pm$ $...$    & 1.48\\
Max. disk        & 0.88	 & $<0.1$      & 936.1 $\pm$ $...$   & 0.89	   &&  0.88	   &  $<0.1$        & 954.8 $\pm$ $...$    & 1.26\\
Model $\Upsilon_{*}^{3.6}$ disk & 0.33 & $<0.1$  & 630.7 $\pm$ $...$     & 0.98	   &&  0.33    &  $<0.1$   & 1143.6 $\pm$ $...$    & 1.37\\
\MLfree\ disk    & 1.15  & $<0.1$      &  665.7 $\pm$ $...$     & 1.26    &&          &               &         &     \\
\hline\\
\noalign{\vskip 2pt}
 & \multicolumn{4}{c}{Pseudo-isothermal halo (entire region)} && \multicolumn{4}{c}{Pseudo-isothermal halo ($<$ 6.0 kpc)}\\
\cline{2-5} \cline{7-10} \\
\multicolumn{1}{c}{\ML\ assumption}    & \multicolumn{1}{c}{$\langle\MLsps\rangle$}   &  \multicolumn{1}{c}{$R_C$} &   \multicolumn{1}{c}{$\rho_0$}    & \multicolumn{1}{c}{$\chi_{red.}^2$}     && \multicolumn{1}{c}{$\langle\MLsps\rangle$} & \multicolumn{1}{c}{$R_C$}      & \multicolumn{1}{c}{$\rho_0$}   & \multicolumn{1}{c}{$\chi_{red.}^2$}\\
\multicolumn{1}{c}{(10)} & \multicolumn{1}{c}{(11)}   & \multicolumn{1}{c}{(12)}      & \multicolumn{1}{c}{(13)}    & \multicolumn{1}{c}{(14)}    && \multicolumn{1}{c}{(15)}  & \multicolumn{1}{c}{(16)} & \multicolumn{1}{c}{(17)}  & \multicolumn{1}{c}{(18)} \\
\hline
Min. disk        & 0.00	  &  1.47 $\pm$ 0.06       & 44.6 $\pm$ 2.2         & 0.16         && 0.00	&   1.49 $\pm$ 0.07       & 44.1 $\pm$ 2.6         & 0.21\\
Min. disk+gas    & 0.00 	  &  1.25 $\pm$ 0.05       & 43.8 $\pm$ 2.4         & 0.13         && 0.00 	&   1.25 $\pm$ 0.06       & 43.7 $\pm$ 2.8         & 0.18\\
Max. disk        & 0.88	  &  1.61 $\pm$ 0.15       & 21.8 $\pm$ 2.5         & 0.25         && 0.88	&   1.62 $\pm$ 0.18       & 21.8 $\pm$ 3.0         & 0.34\\
Model $\Upsilon_{*}^{3.6}$ disk & 0.33  &  1.36 $\pm$ 0.07 & 34.8 $\pm$ 2.4         & 0.17         && 0.33  &    1.36 $\pm$ 0.09       & 34.7 $\pm$ 2.9         & 0.23\\
\MLfree\ disk    & $-0.99$ & 0.98 $\pm$ 0.05       & 87.1 $\pm$ 14.0        & 0.14         &&       &                 &     \\
\hline
\end{tabular}
\medskip\noindent
\begin{minipage}{156mm}
\noindent
\\
{\bf Note.$\--$}
{\bf (1)(10):} The stellar mass-to-light ratio \ML\ assumptions. ``Model \MLsps\ disk'' uses the values derived from the population synthesis models in Section 4.2.
\MLfree\ has \MLsps\ as a free parameter.
{\bf (2)(6)(11)(15):} Average \ML\ in the 3.6 $\mu$m band (\Msun/\Lsun).
{\bf (3)(7):} Concentration parameter c of NFW halo model (NFW 1996, 1997).
{\bf (4)(8):} The rotation velocity (\kms)\,at radius $R_{200}$ where the density constrast exceeds 200 (Navarro \etal\ 1996).
{\bf (5)(9)(14)(18):} Reduced $\chi^{2}$ value.
{\bf (12)(16):} Fitted core-radius of pseudo-isothermal halo model (kpc).
{\bf (13)(17):} Fitted core-density of pseudo-isothermal halo model ($10^{-3}$ \cubedens).
{\bf ($...$):} blank due to unphysically large value or not well-constrained uncertainties.
\end{minipage}
\end{center}
\end{table*}

\begin{figure*}
\epsscale{1.3}
\includegraphics[angle=0,width=1.0\textwidth,bb=10 325 540 755,clip=]{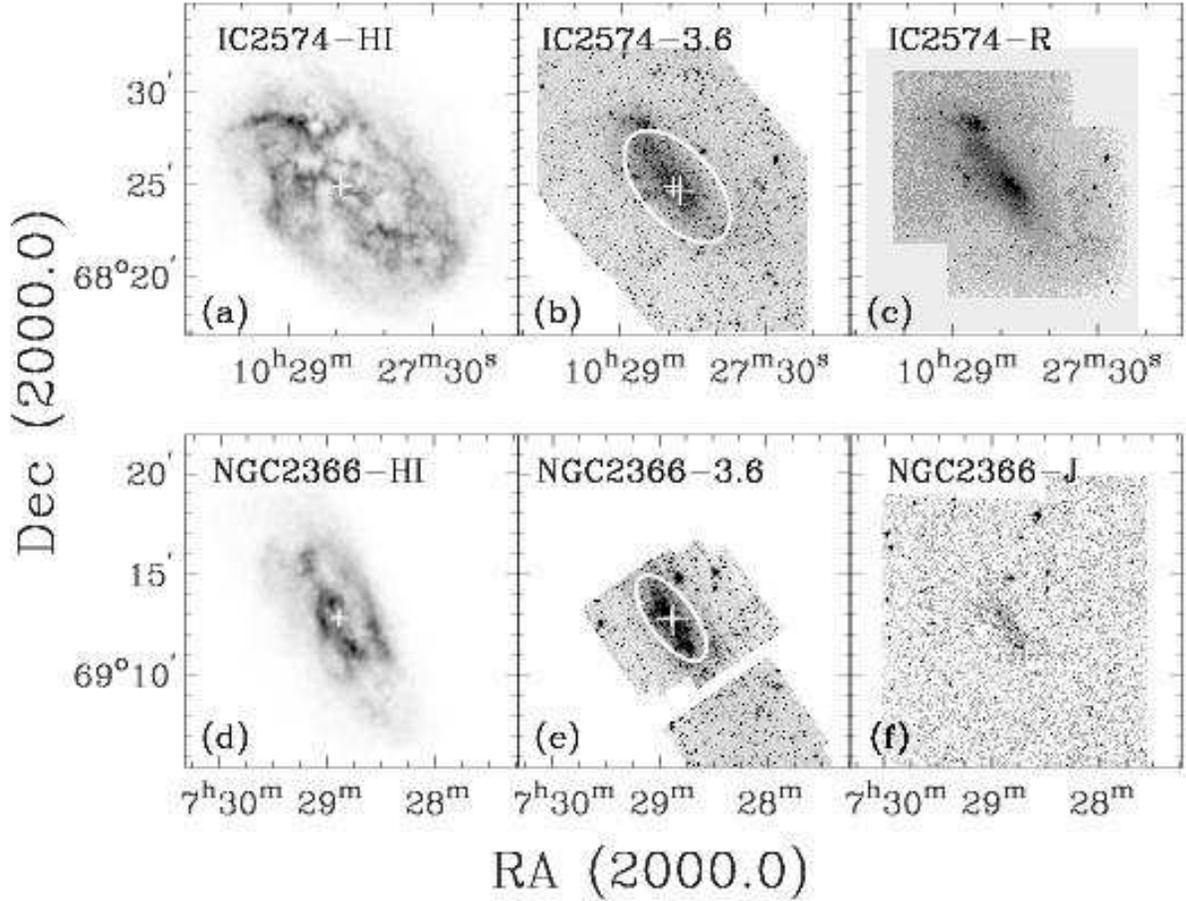}
\caption{Total intensity maps of IC 2574 and NGC 2366 in various bands.
  (a)(d): Integrated THINGS H{\sc i} maps. The crosses indicate the
  derived dynamical centers in this paper.  (b)(e): 3.6 $\mu$m image
  with superimposed the photometric centers as listed in NED (large
  crosses) and the dynamical centers as derived in this paper from the
  bulk velocity fields (small crosses).  The ellipses indicate the
  ellipse fit discussed in Section 3.1.  (c): Optical $R$-band image
  obtained with the 2.1m telescope at KPNO as part of the
  ancillary SINGS data (Kennicutt \etal\ 2003).  (f): $J$-band image
  from 2MASS.
\label{FIGURE1}}
\end{figure*}

\begin{figure}
\epsscale{1.0}
\includegraphics[angle=0,width=1.0\textwidth,bb=5 180 565 740,clip=]{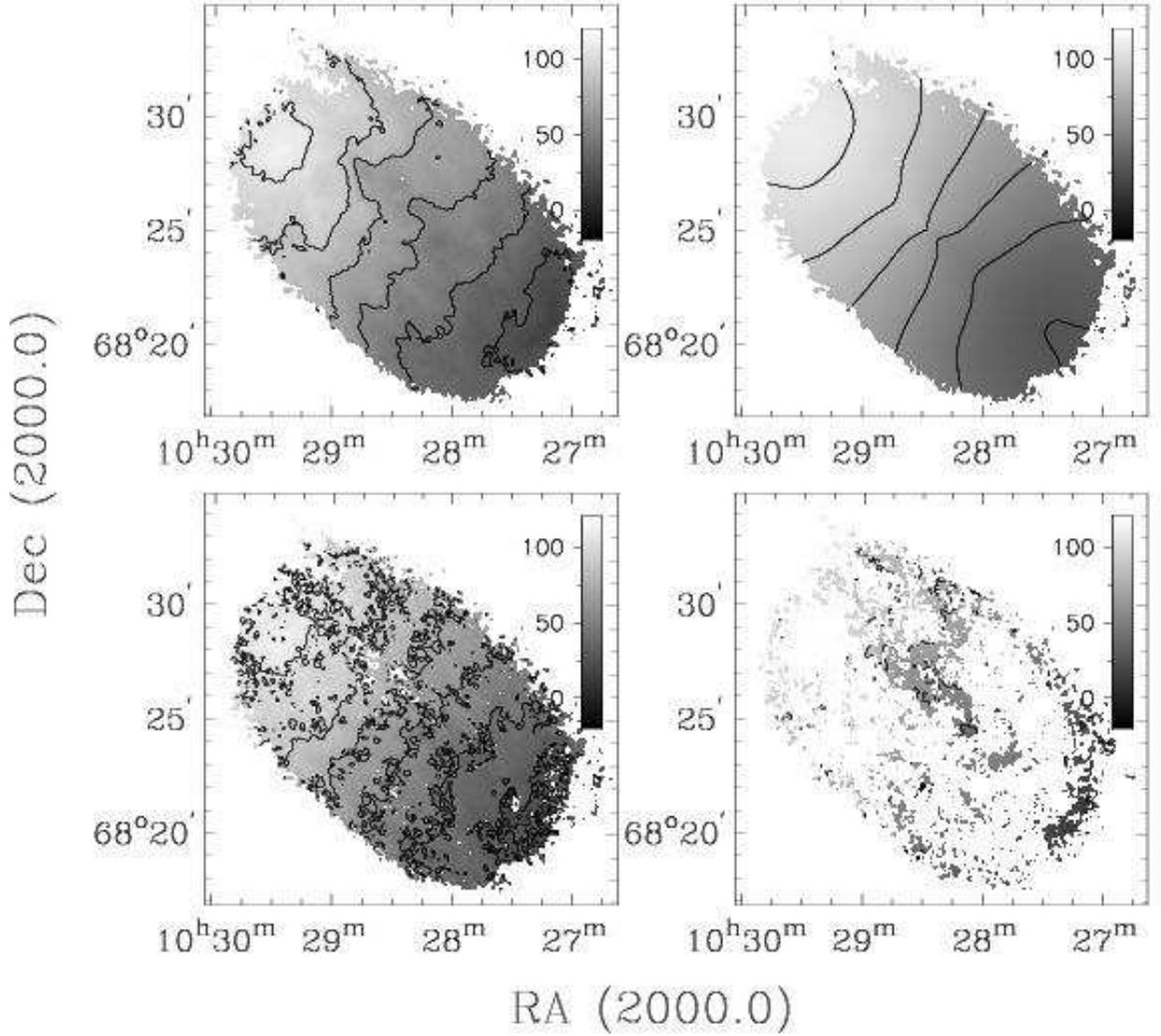}
\caption{The velocity fields of IC 2574. 
Top-left: Intensity-weighted mean (IWM) velocity field from THINGS.
Top-right: Initial value velocity field for the bulk motion.
Bottom-left: Bulk velocity field.
Bottom-right: Velocity field of the non-circular motions.
Velocity contours run from $-60$\,\kms\,to $120$\,\kms\,with a
spacing of $20$\,\kms.
Note that the strongest non-circular motions seen in the central region 
of the IWM velocity field are removed in the extracted bulk velocity field. 
See Section 3.1 for a full description.
\label{FIGURE2}}
\end{figure}

{\clearpage}

\begin{figure}
\epsscale{1.0}
\includegraphics[angle=0,width=1.0\textwidth,bb=5 180 565 740,clip=]{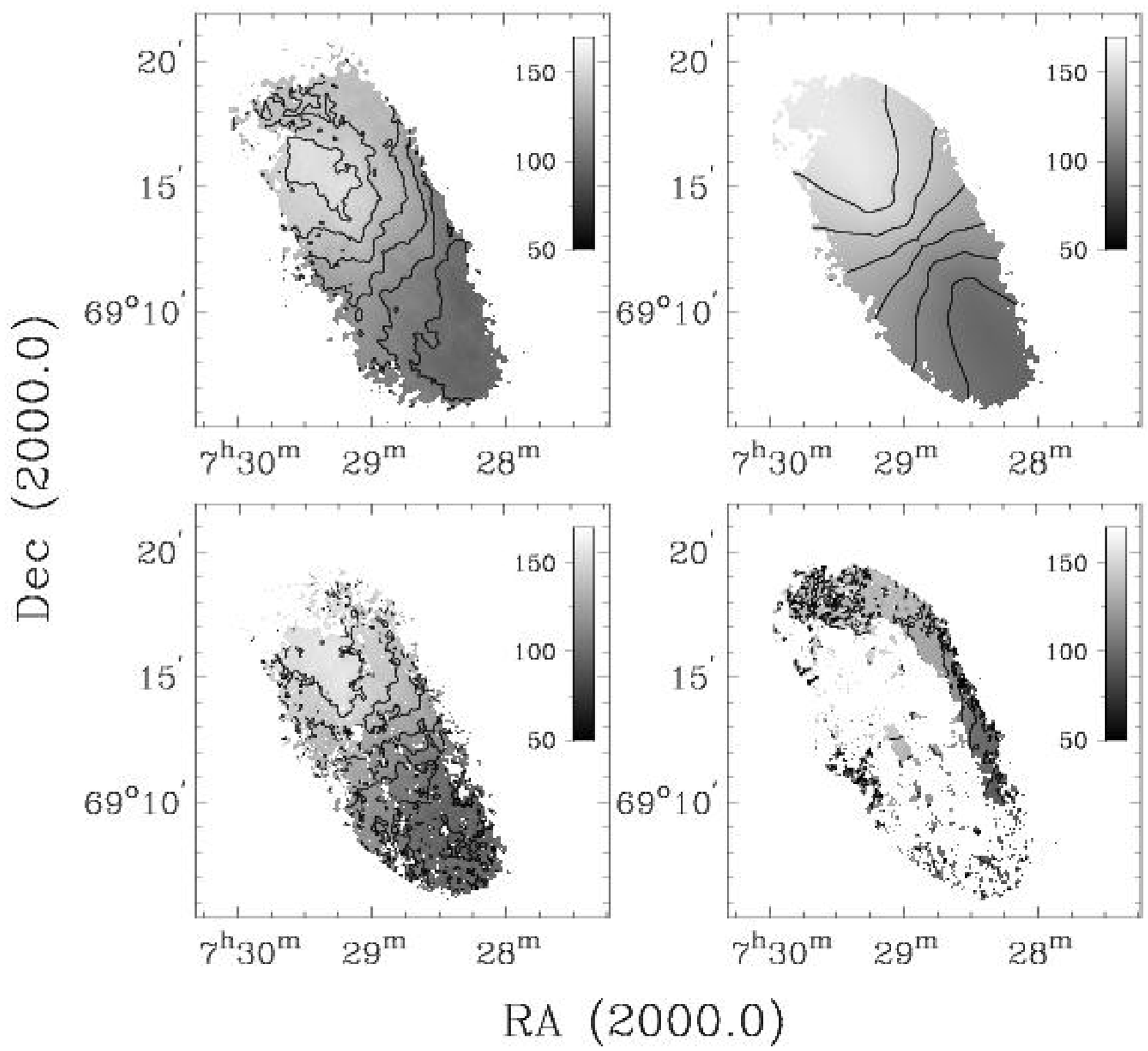}
\caption{The velocity fields of NGC 2366. 
Top-left: Intensity-weighted mean (IWM) velocity field from THINGS.
Top-right: Initial value velocity field for the bulk motion.
Bottom-left: Bulk velocity field.
Bottom-right: Velocity field of the non-circular motions.
Velocity contours run from $-20$\,\kms\,to $170$\,\kms\,with a
spacing of $15$\,\kms.
Note that the strongest non-circular motions are found in the outer part.
See Section 3.1 for a full description.
\label{FIGURE3}}
\end{figure}
 {\clearpage}

\begin{figure}
\epsscale{1.0}
\includegraphics[angle=0,width=1.0\textwidth,bb=24 490 564 755,clip=]{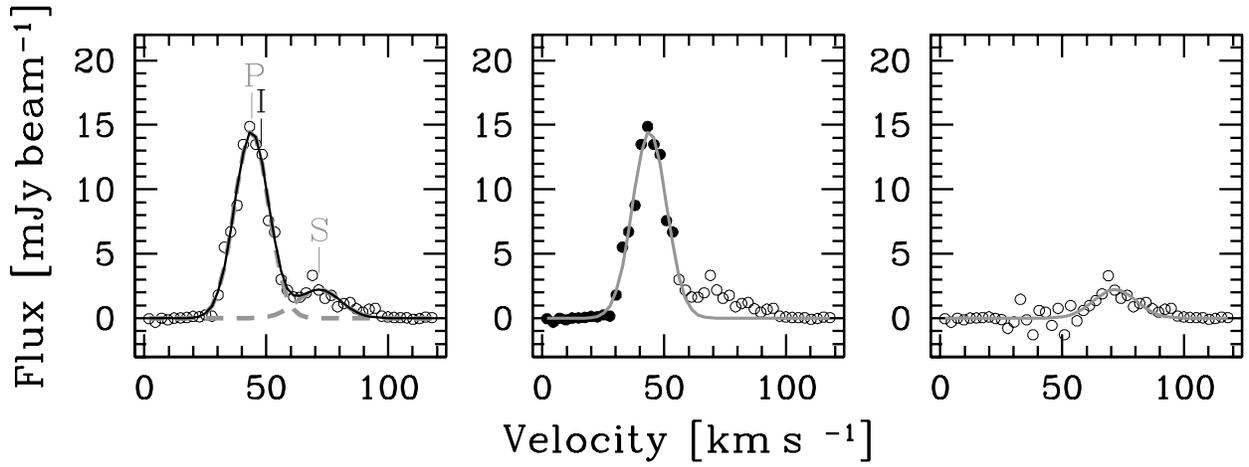}
\caption{A schematic example of Gaussian decomposition using two Gaussian functions. 
Left: Decomposed profiles with two Gaussian functions. 
P and S represent the central values of the primary and secondary Gaussian components 
and I denotes the IWM value. 
Middle: The primary Gaussian component is fitted only to the data points that are
unaffected by the secondary component (filled circles). 
Right: The secondary Gaussian component is fitted to the residual after subtracting 
the primary Gaussian component from the raw profile. 
Note that the IWM value will deviate from that of the primary
component by an amount which depends on the significance of the secondary.
Section 3.1 gives a more detailed description. 
\label{FIGURE4}}
\end{figure}
 {\clearpage}

\begin{figure}
\epsscale{0.6}
\plotone{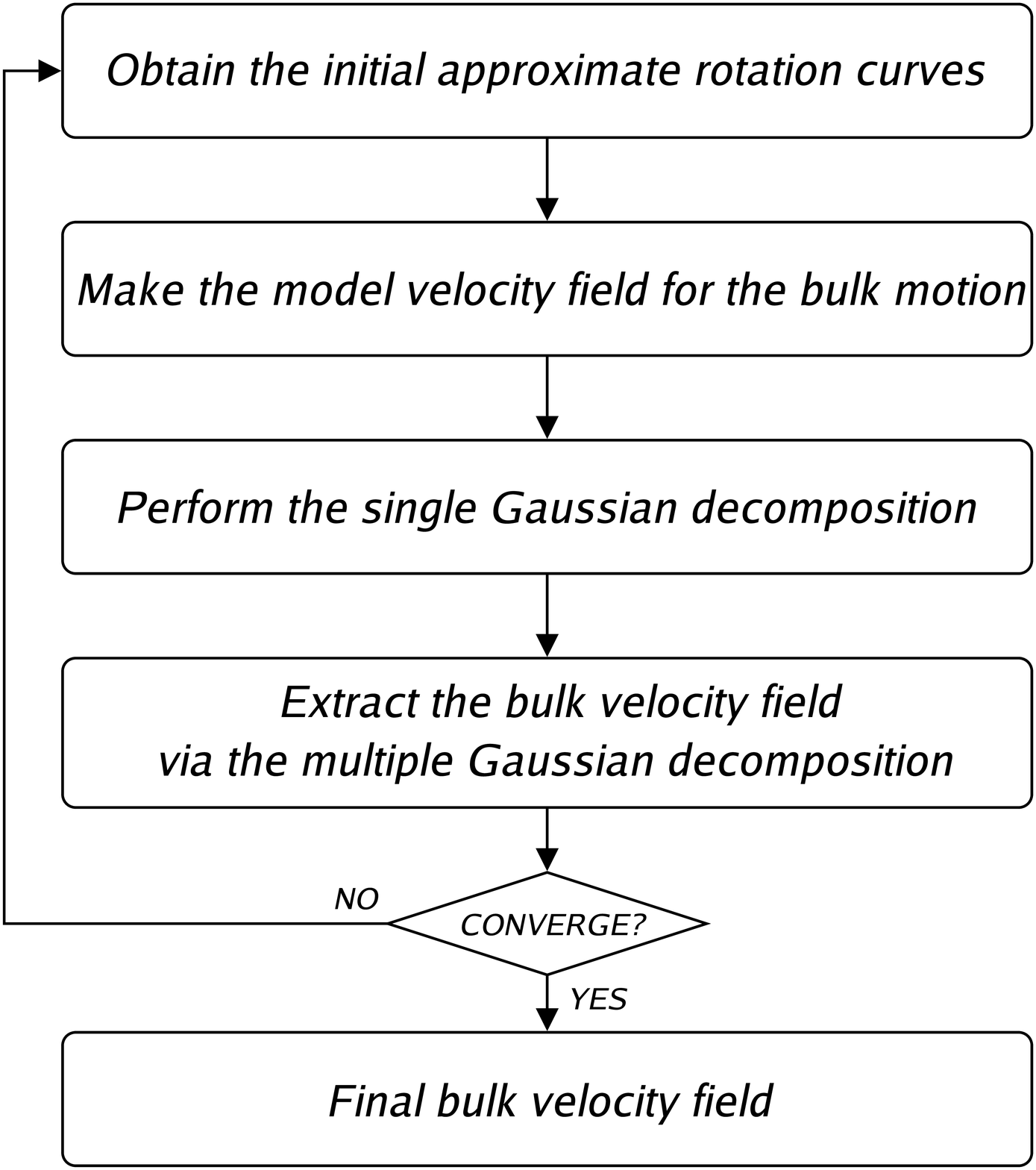}
\caption{A schematic diagram of the bulk-motion extraction method. This flow chart shows the overall procedure
for extracting the bulk velocity field from the H{\sc i} data cube. The procedure is iterated until the
convergence criterion is met (see text for details). The method is fully described in Section 3.1.
\label{FIGURE5}}
\end{figure}
 {\clearpage}

\begin{figure}
\epsscale{1.0}
\includegraphics[angle=0,width=1.0\textwidth,bb=20 175 605 730,clip=]{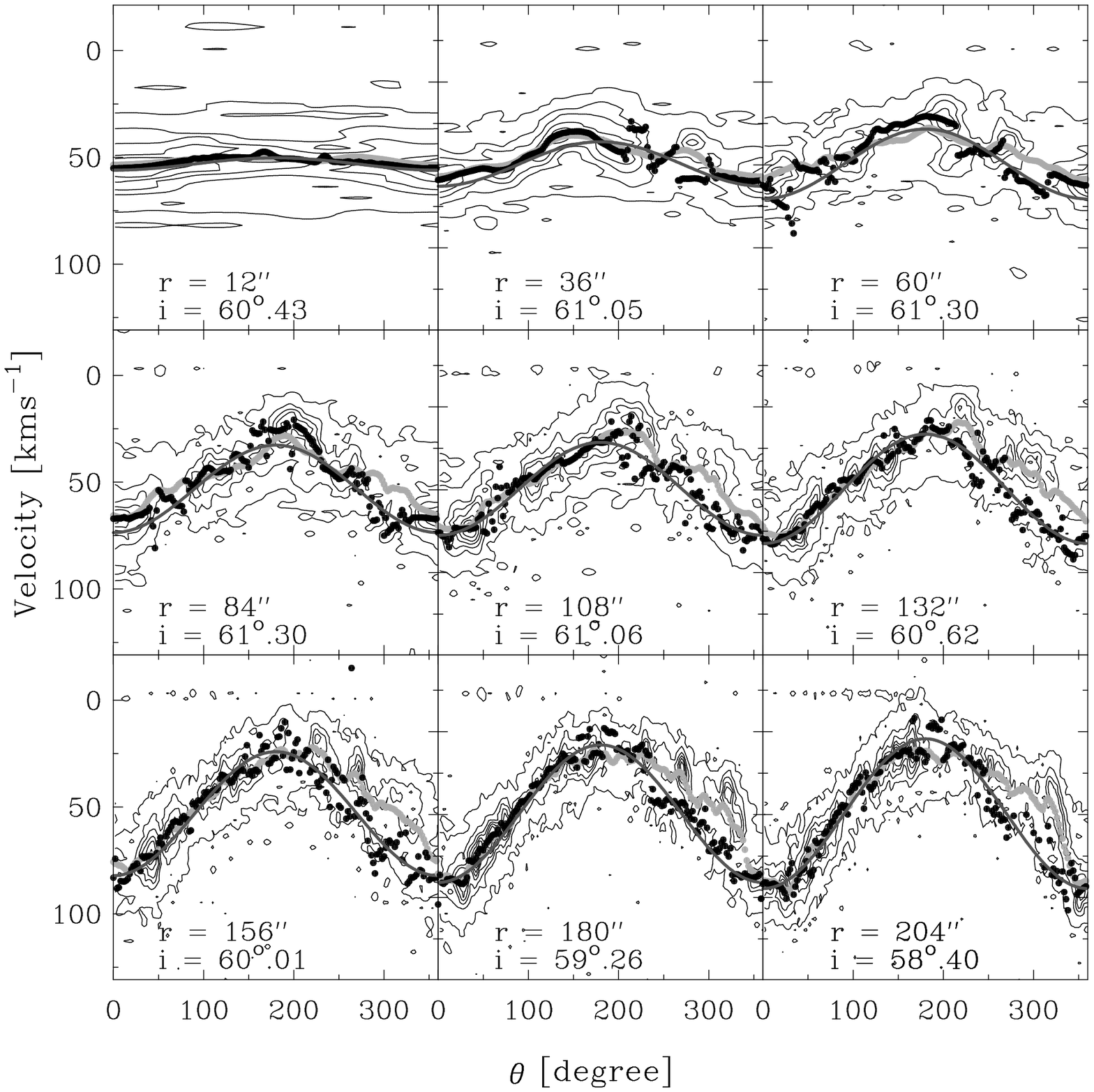}
\caption{Azimuthal position-velocity diagrams of IC 2574. The observed velocities are displayed as a function 
of $\theta$ along ellipses in the plane of sky. The angle $\theta$ is measured anticlockwise with respect to
the receding major axis in the plane of the sky. Each ellipse is defined by a radius and inclination
as labeled in each panel. All ellipses have identical dynamical centers as derived in this paper from
the bulk velocity field (see Fig. 1). 
The width of the ellipse rings used for the azimuthal integration is $1.5\arcsec$.
Contours start at +1$\sigma$ in steps of +5$\sigma$ with $\sigma$=0.56 \mJy. The light-gray dots represent the IWM velocity
and the black dots indicate the bulk velocity derived by our method in Section 3.3. The dark-gray 
solid line on each panel represents
the expected ``bulk'' motion. In many cases, especially for $R < 300\arcsec$ and $\theta \sim 320^{\circ}$, the IWM
velocities deviate significantly from the symmetric shape of the rotation curve. 
\label{FIGURE6A}}
\end{figure}
 {\clearpage}

\begin{figure}
\figurenum{6}
\epsscale{1.0}
\includegraphics[angle=0,width=1.0\textwidth,bb=20 175 605 730,clip=]{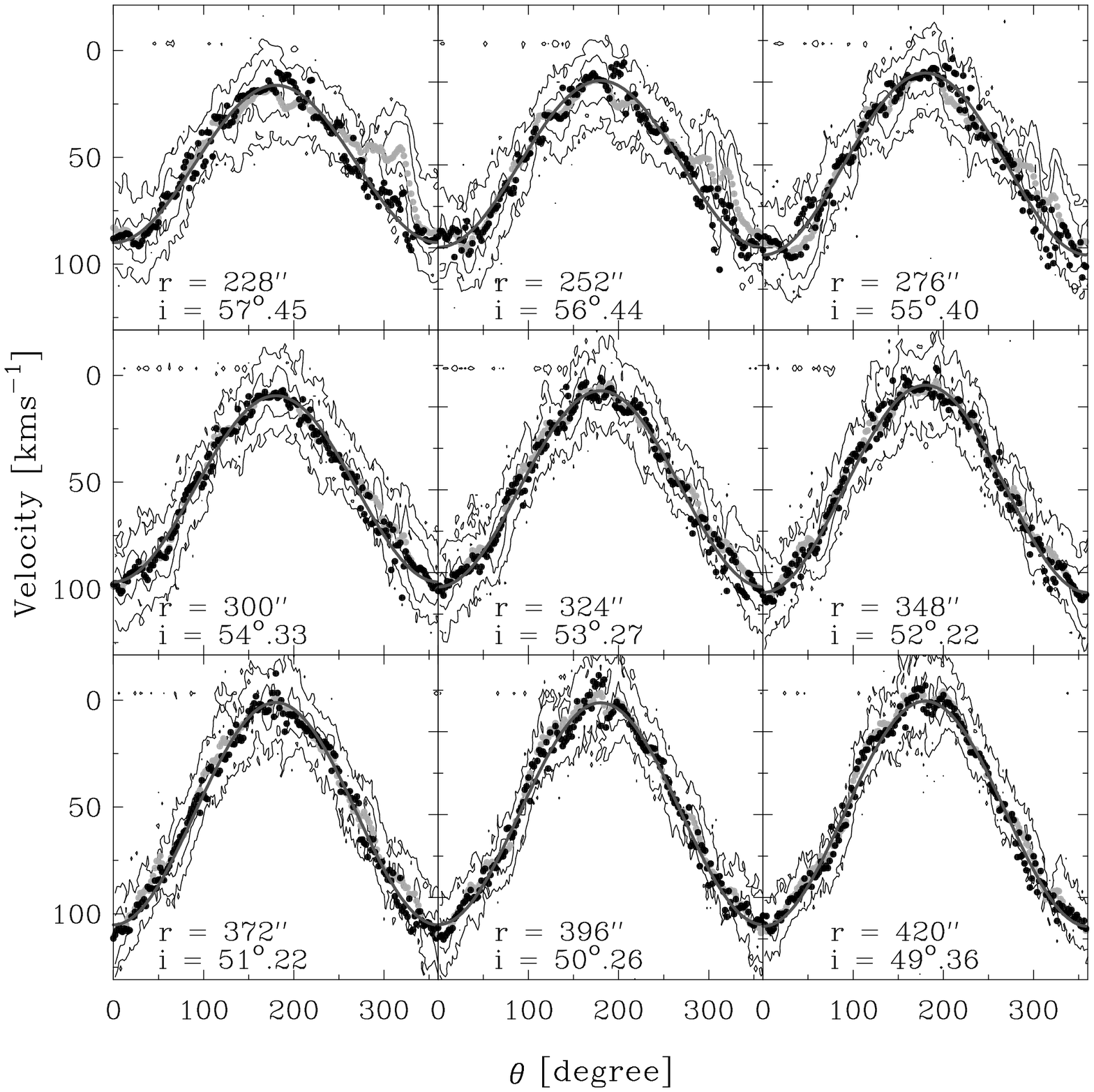}
\caption{Position-velocity diagrams of IC 2574 (cont'd).
\label{FIGURE6B}}
\end{figure}
 {\clearpage}

\begin{figure}
\epsscale{1.0}
\includegraphics[angle=0,width=1.0\textwidth,bb=20 175 605 730,clip=]{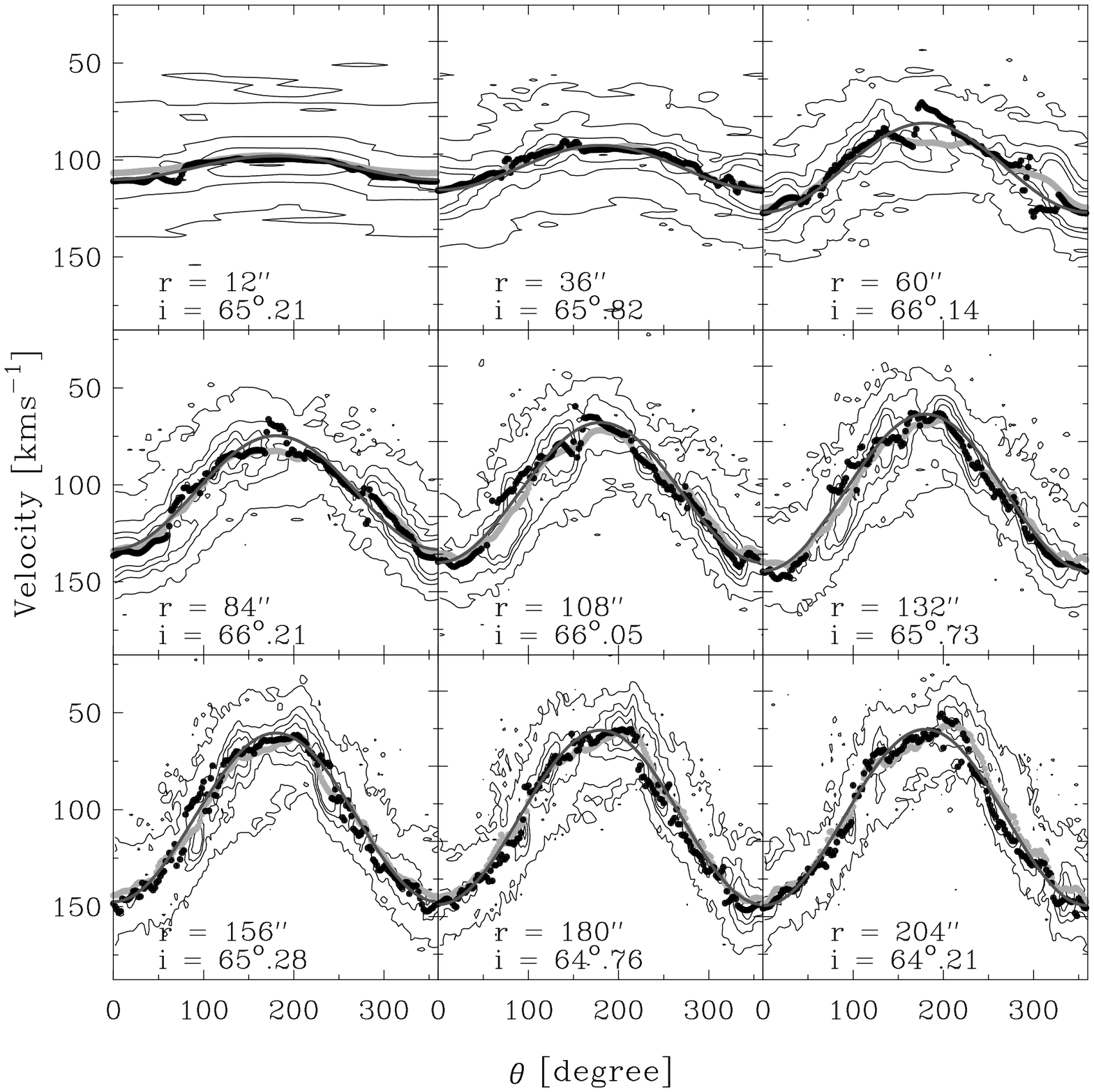}
\caption{Azimuthal position-velocity diagrams of NGC 2366. The observed velocities are displayed as a function 
of $\theta$ along ellipses in the plane of sky. The angle $\theta$ is measured anticlockwise with respect to
the receding major axis in the plane of the sky. Each ellipse is defined with a radius and inclination
as labeled in each panel. All ellipses have identical dynamical centers as derived in this paper from 
the bulk velocity field (see Fig. 1). 
The width of the ellipse rings used for the azimuthal integration is $1.5\arcsec$.
Contours start at +2$\sigma$ in steps of +10$\sigma$ with $\sigma$=0.52 \mJy.
The light-gray dots represent the IWM velocity
and the black dots indicate the bulk velocity derived by our method in Section 3.3.
The dark-gray solid line on each panel represents the expected ``bulk'' motion. 
In many cases, especially for $R > 300\arcsec$ and $\theta \sim 270^{\circ}$, the IWM
 velocities deviate significantly from the symmetric shape of the rotation curve. 
\label{FIGURE7A}}
\end{figure}
 {\clearpage}

\begin{figure}
\epsscale{1.0}
\figurenum{7}
\includegraphics[angle=0,width=1.0\textwidth,bb=20 175 605 730,clip=]{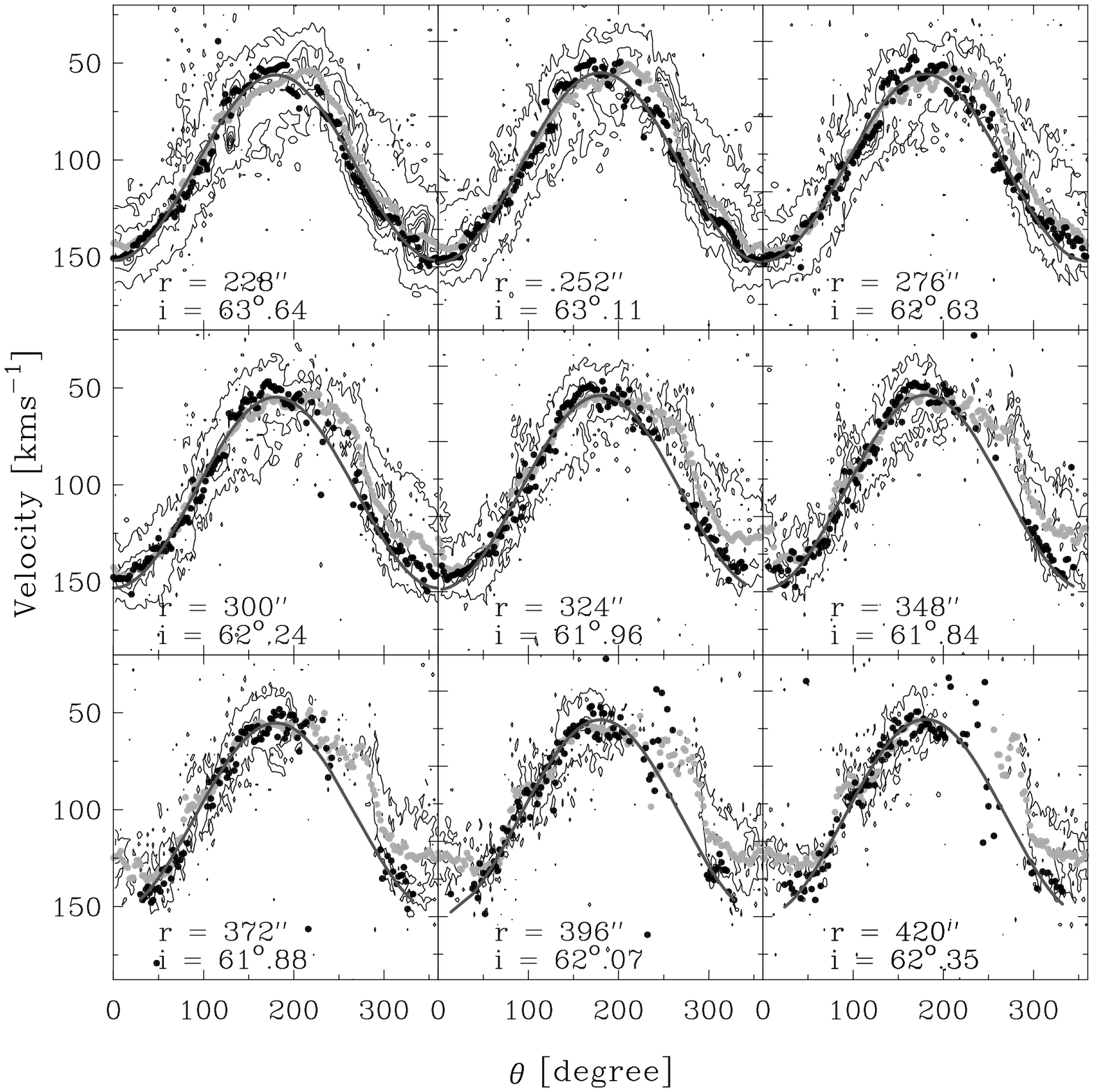}
\caption{Position-velocity diagrams of NGC 2366 (cont'd).
\label{FIGURE7B}}
\end{figure}
 {\clearpage}

\begin{figure}
\epsscale{1.0}
\plotone{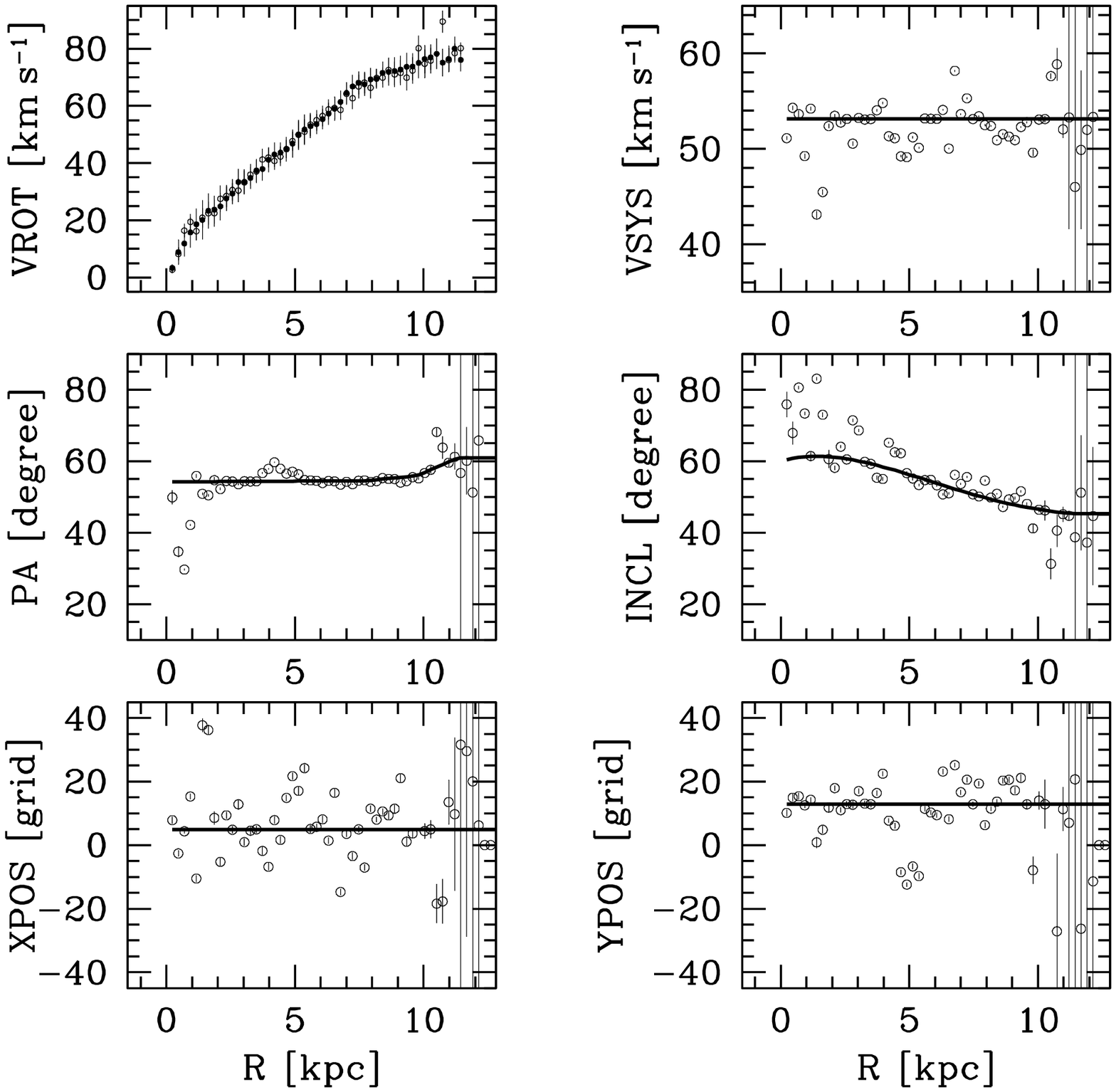}
\caption{The tilted ring model derived from the bulk velocity field of IC 2574. The open
black circles in all panels indicate the fit made with all parameters free. 
The filled black circles in the VROT panel were derived using the entire velocity field
after fixing other ring parameters to the values (solid lines) as shown in the panels. 
See Fig. 11 for a larger scale.
In all other panels the solid lines show the values as a function
of radius adopted for the final tilted ring model.
\label{FIGURE8}}
\end{figure}
 {\clearpage}

\begin{figure}
\epsscale{1.0}
\plotone{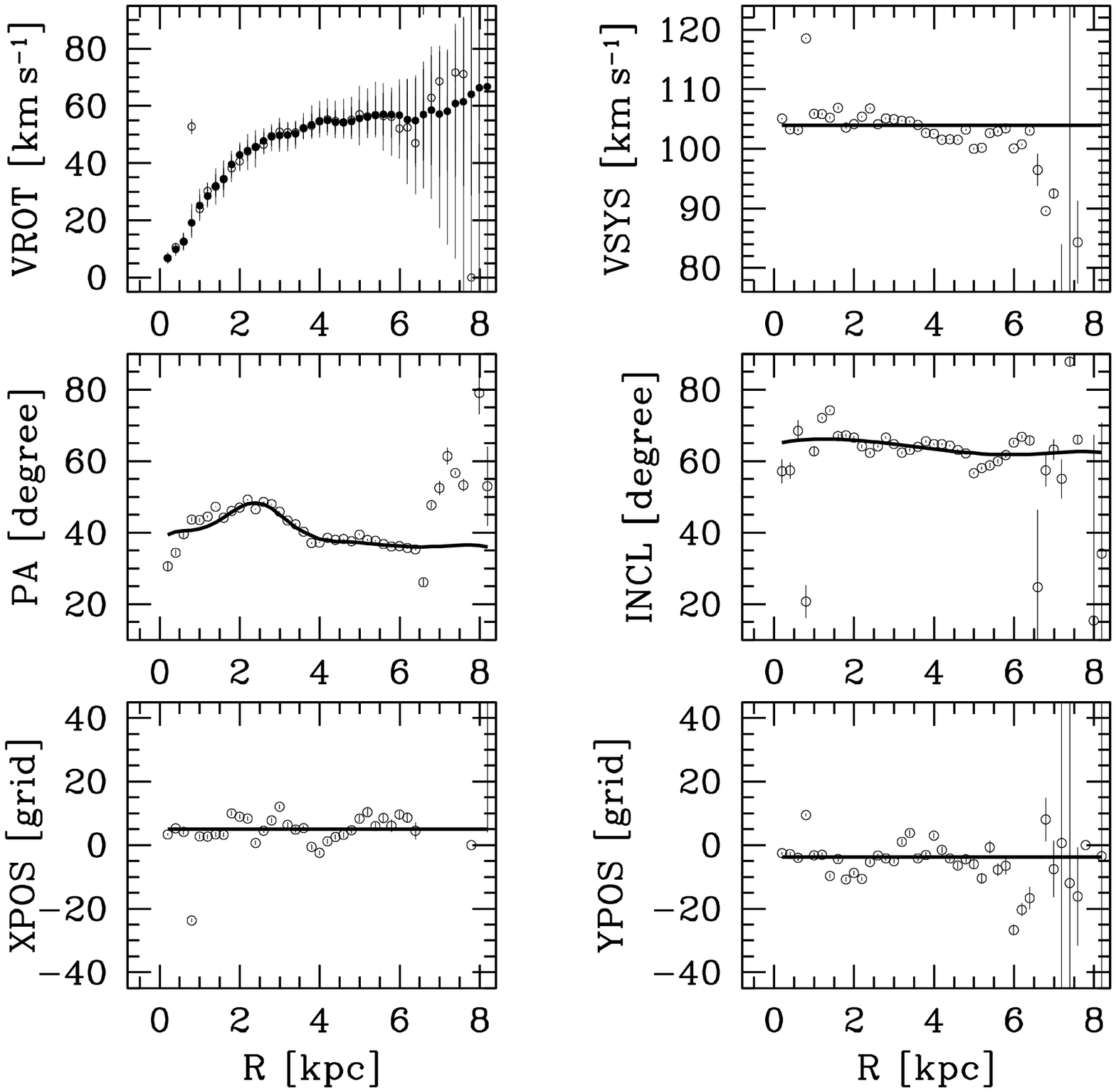}
\caption{The tilted ring model derived from the bulk velocity field of NGC 2366. The open
black circles in all panels indicate the fit made with all ring parameters free. 
The filled black circles in the VROT panel were derived using the entire velocity field
after fixing other ring parameters to the values (solid lines) as shown in the panels. 
See Fig. 13 for a larger scale.
In all other panels the solid lines show the values as a function
of radius adopted for the final tilted ring model.
\label{FIGURE9}}
\end{figure}
 {\clearpage}

\begin{figure}
\epsscale{1.0}
\plotone{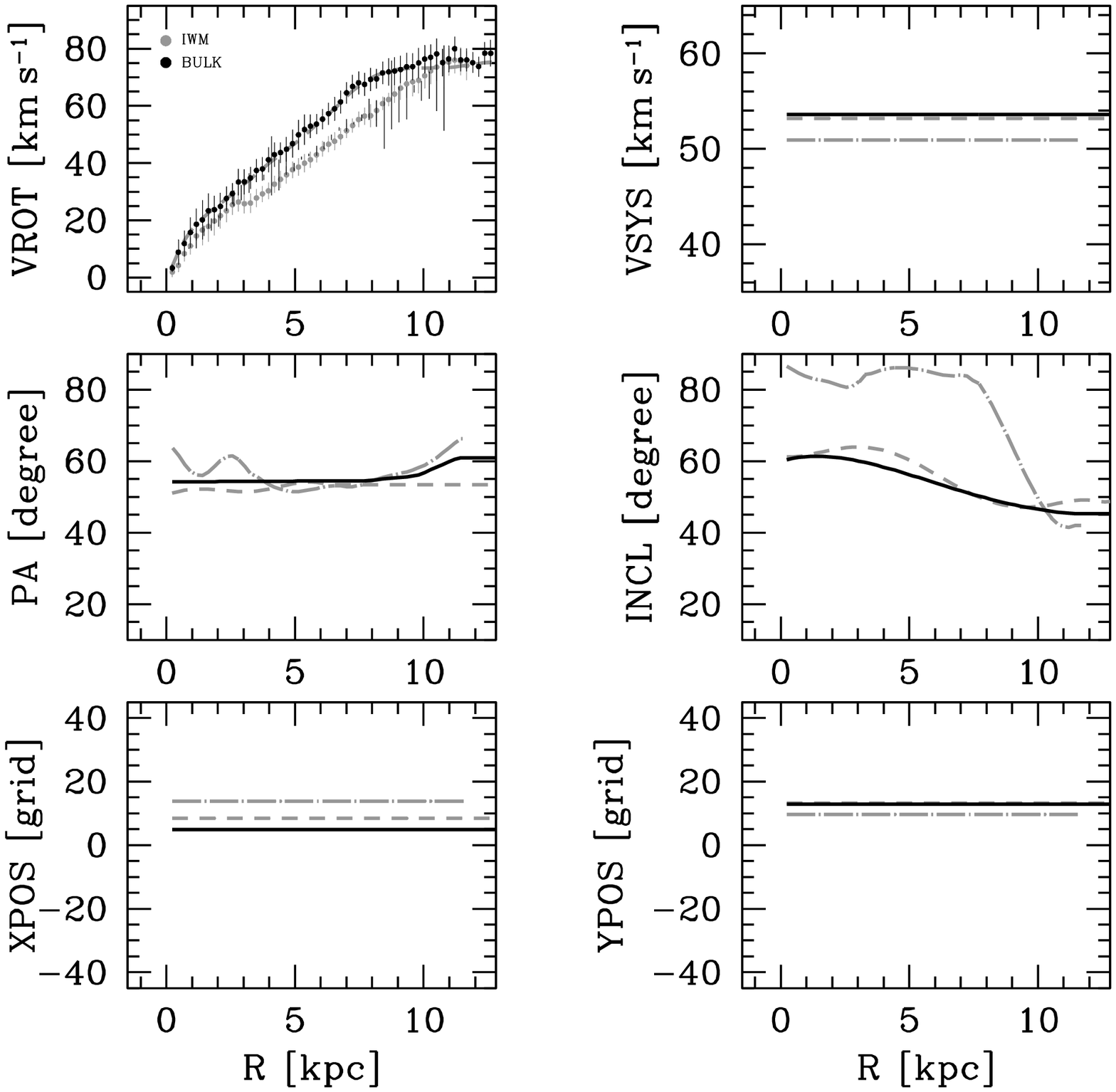}
\caption{Comparison of the H{\sc i} rotation curves derived from the IWM and bulk velocity fields of IC 2574. 
The filled gray circles and long dash-dotted lines represent the rotation curves from the IWM 
velocity field. The gray dashed lines are used as initial condition for a bulk velocity field model.
The black dots and solid lines show the adopted rotation curves of IC 2574 
using the bulk velocity field. The large difference in inclinations between IWM and bulk 
velocity fields is clearly evident in the panel of INCL (inclination) and this results 
in a significant difference ($\sim$14\,\kms) of VROT (rotation velocity). 
More details are given in Section 3.3.
\label{FIGURE10}}
\end{figure}
 {\clearpage}

\begin{figure}
\epsscale{1.0}
\plotone{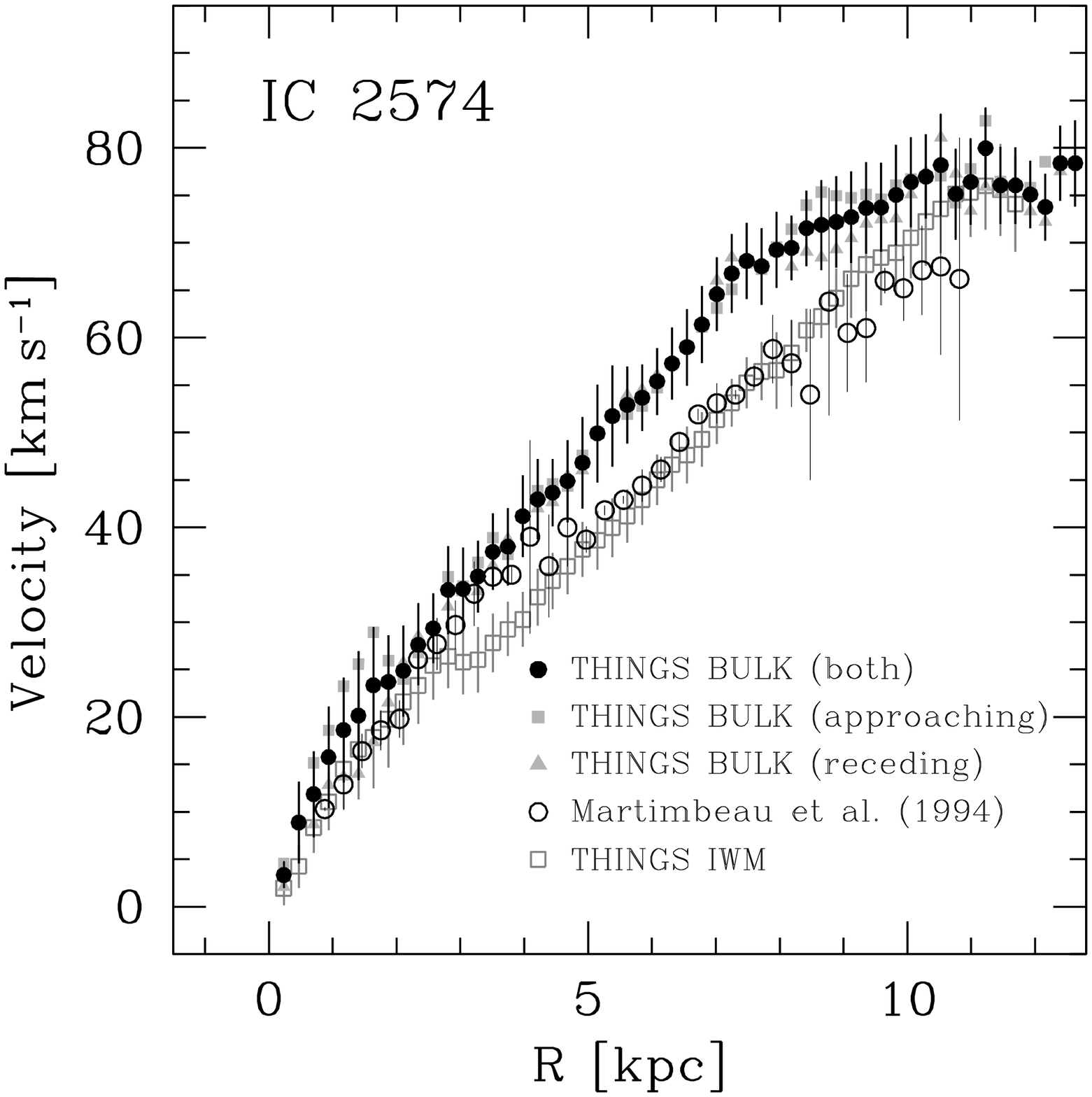}
\caption{Comparison of the rotation curve of IC 2574 with the rotation curves from literature.
The filled black circles and open squares represent the rotation curves from the bulk and 
IWM velocity field, respectively.
The filled gray squares and triangles are derived using 
only the approaching and receding sides of the bulk velocity field with the final tilted ring model
in Fig. 8 (black solid curves).
Open circles were adopted from Martimbeau \etal\ (1994).
More details are given in Section 3.3.
\label{FIGURE11}}
\end{figure}
 {\clearpage}

\begin{figure}
\epsscale{1.0}
\plotone{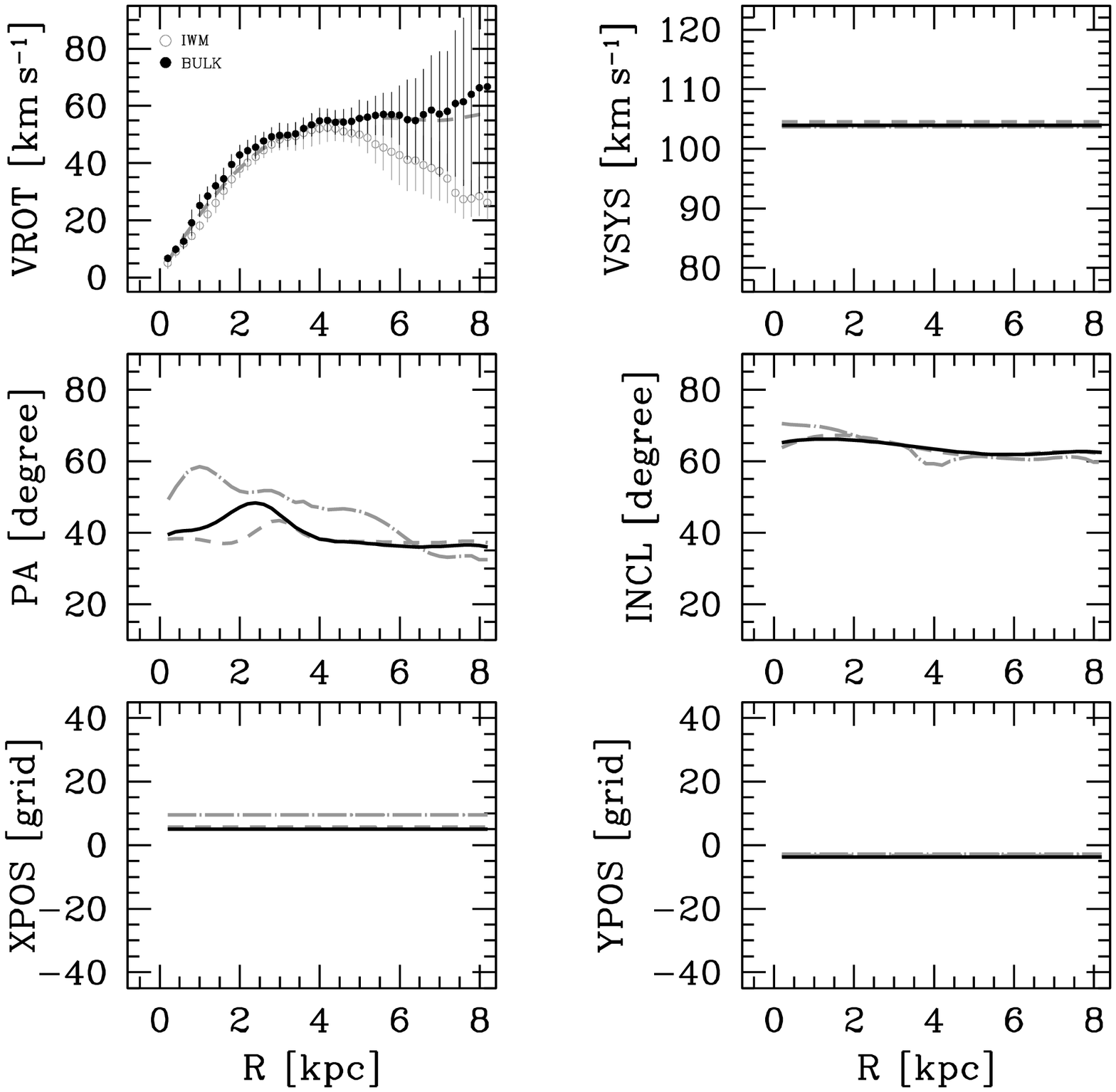}
\caption{Comparison of the H{\sc i} rotation curves derived from the IWM and bulk velocity fields of NGC 2366. 
The open gray circles and long dash-dotted lines represent the rotation curves from the IWM 
velocity field. The gray dashed lines are used as initial condition for a bulk velocity field model.
The black dots and solid lines show the adopted rotation curves of NGC 2366 using the bulk velocity field. 
See Fig. 13 for a detailed comparison with literature values.
\label{FIGURE12}}
\end{figure}
 {\clearpage}

\begin{figure}
\epsscale{1.0}
\plotone{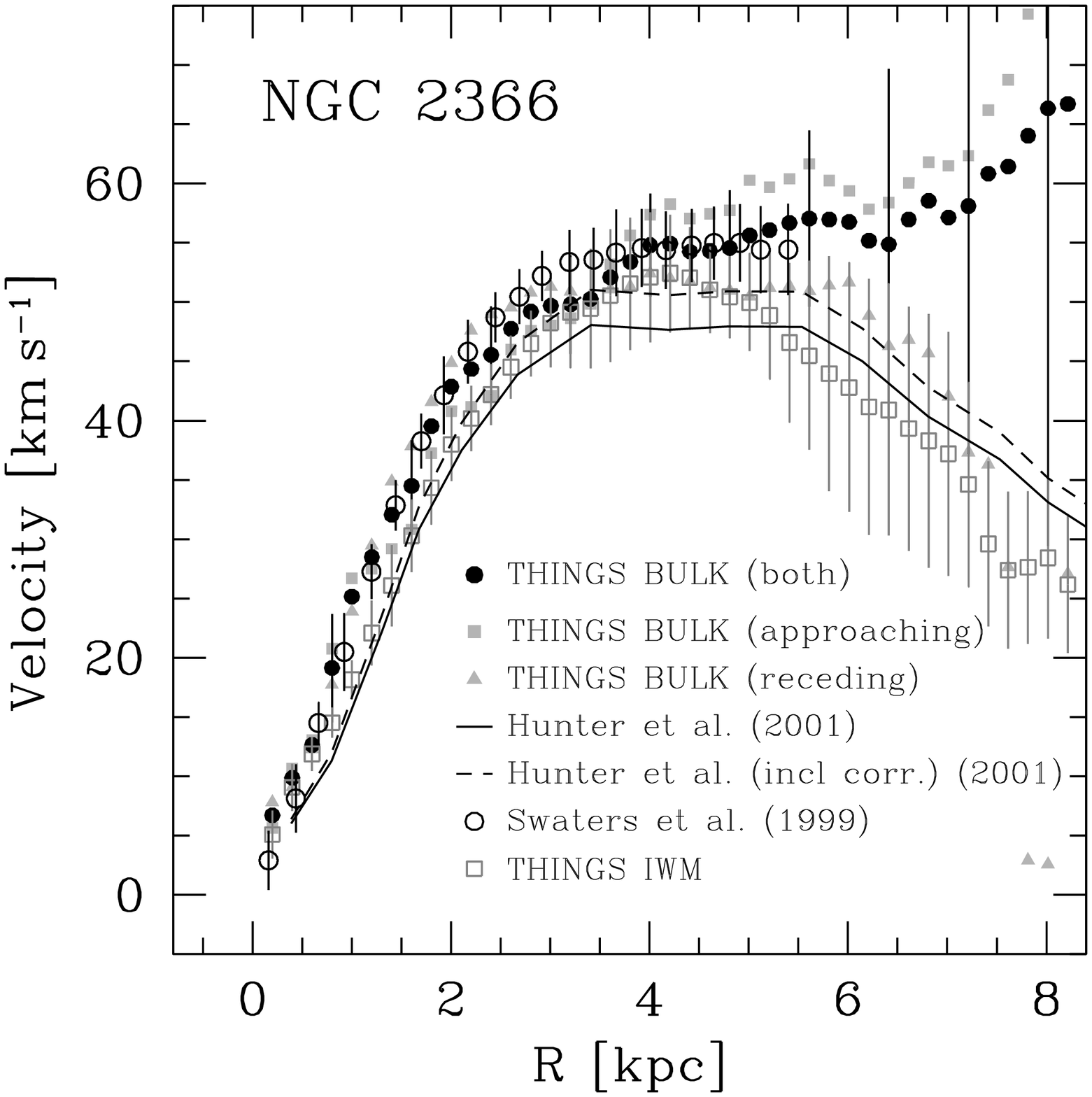}
\caption{Comparison of the rotation curve of NGC 2366 with rotation curves from the literature.
The filled black circles and open squares represent the rotation curves from the bulk and 
IWM velocity field, respectively. 
The filled gray squares and triangles are derived using 
only the approaching and receding sides of the bulk velocity field with the final tilted ring model
in Fig. 9 (black solid curves).
Open circles were adopted from Swaters \etal\ (1999) and
the solid line shows the rotation curve from Hunter \etal\ (2001). The dashed line indicates the
corrected Hunter \etal\ (2001) curve using the inclination of the THINGS IWM curve.
A full description is given in Section 3.3.
\label{FIGURE13}}
\end{figure}
{\clearpage}

\begin{figure}
\epsscale{1.0}
\includegraphics[angle=0,width=0.85\textwidth,bb=19 160 430 697,clip=]{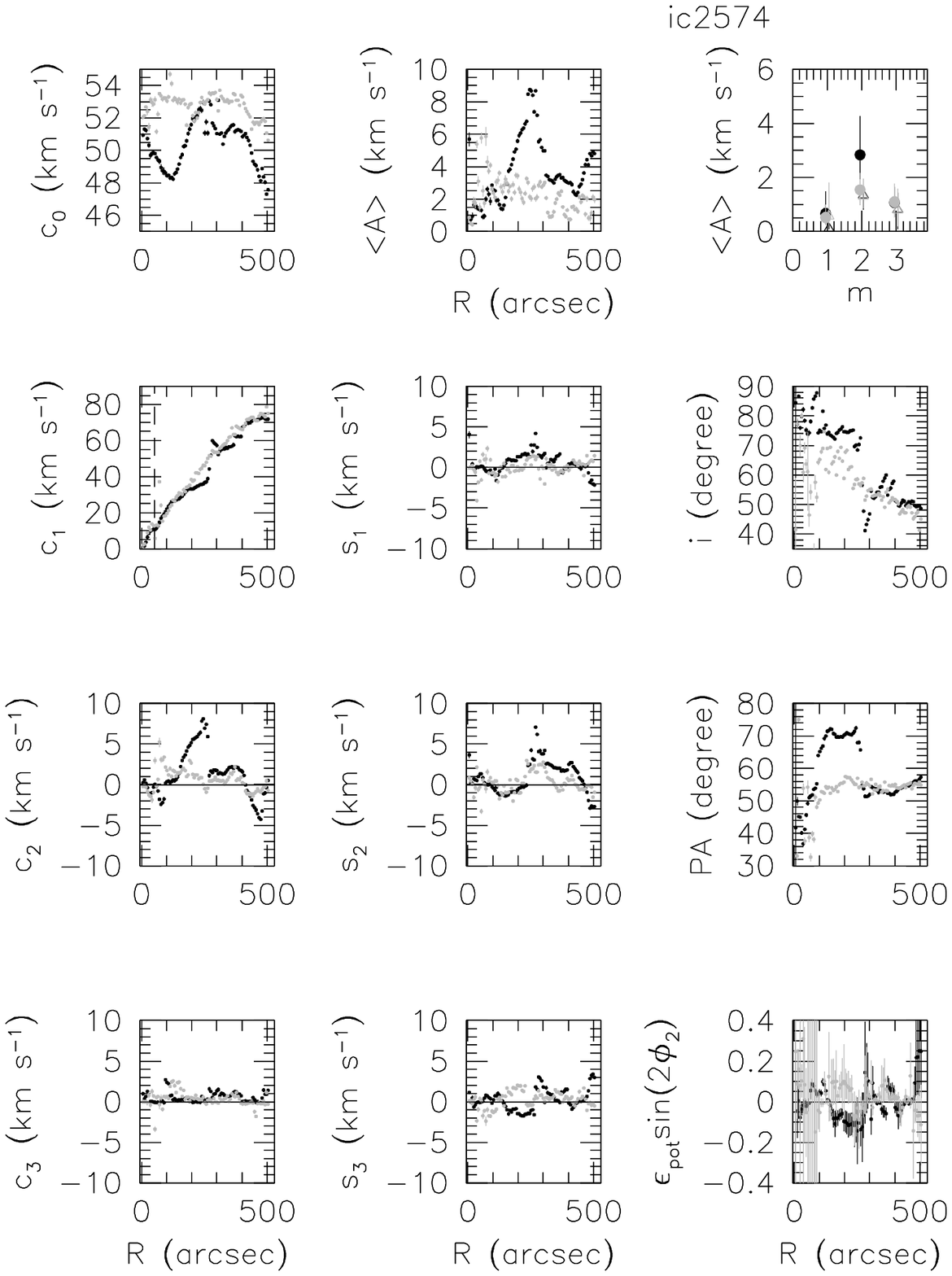}
\caption{Harmonic expansion for IC 2574. 
$c_0$ and $c_1$ are the systemic and the rotation velocities, respectively.
$c_2,c_3,s_1,s_2$, and $s_3$ components quantify non-circular motions.
$\langle A \rangle$ is the median absolute amplitudes as given in Eq. 2, 3, and 4.
Black and gray dots represent the harmonic decomposition
results of the IWM velocity field and the bulk velocity, respectively.
See Section 3.3 for a detailed description.
\label{FIGURE14}}
\end{figure}
 {\clearpage}

\begin{figure}
\epsscale{1.0}
\includegraphics[angle=0,width=1.0\textwidth,bb=5 142 625 688,clip=]{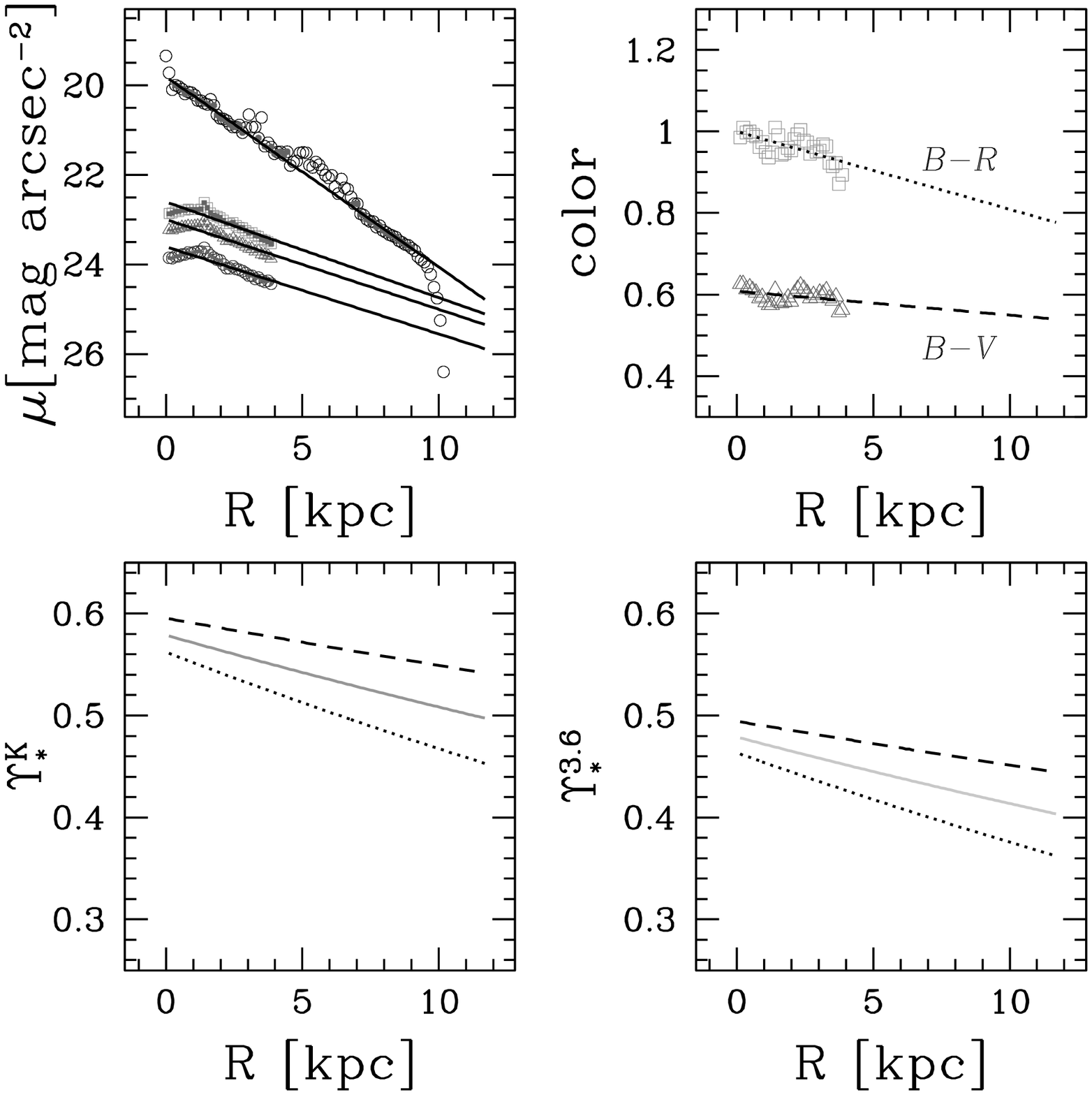}
\caption{Top-left: Azimuthally averaged surface brightness 
profiles (not corrected for inclination) of IC 2574 in the 
3.6 $\mu$m, $R$, $V$, and $B$ bands (top to bottom) derived 
assuming the tilted-ring parameters shown in Fig. 8. 
The curves shown are fitted to the data at $R < 8$ kpc for 3.6 $\mu$m and at
$R < 4$ kpc for $B$, $V$, and $R$ bands (partly filled points).
Top-right: Radial color profiles of IC 2574. 
Dotted line and dashed line are for fits to the $B-R$ and $B-V$ colors, respectively. 
Bottom-left: Radial variation of the model \ML\ in the $K$-band of IC 2574. The dotted 
and dashed lines are computed using optical colors ($B-R$ and $B-V$) and the mean 
value (solid line) is adopted as the final model \ML\ in the $K$-band. The relationships between 
\MLk\ and optical colors (e.g., $B-R$, $B-V$) are adopted from the models of Bell \& de Jong (2001). 
Bottom-right: model \ML\ in the 3.6 $\mu$m band of IC 2574. Eq. 6 is used for converting 
\MLk\ to \MLsps\ (see Section 4.2 for details). The average values of \MLsps\ from different
\ML\ assumptions for IC 2574 are presented in Table 3.
\label{FIGURE15}}
\end{figure}
 {\clearpage}

\begin{figure}
\epsscale{1.0}
\includegraphics[angle=0,width=1.0\textwidth,bb=5 142 625 688,clip=]{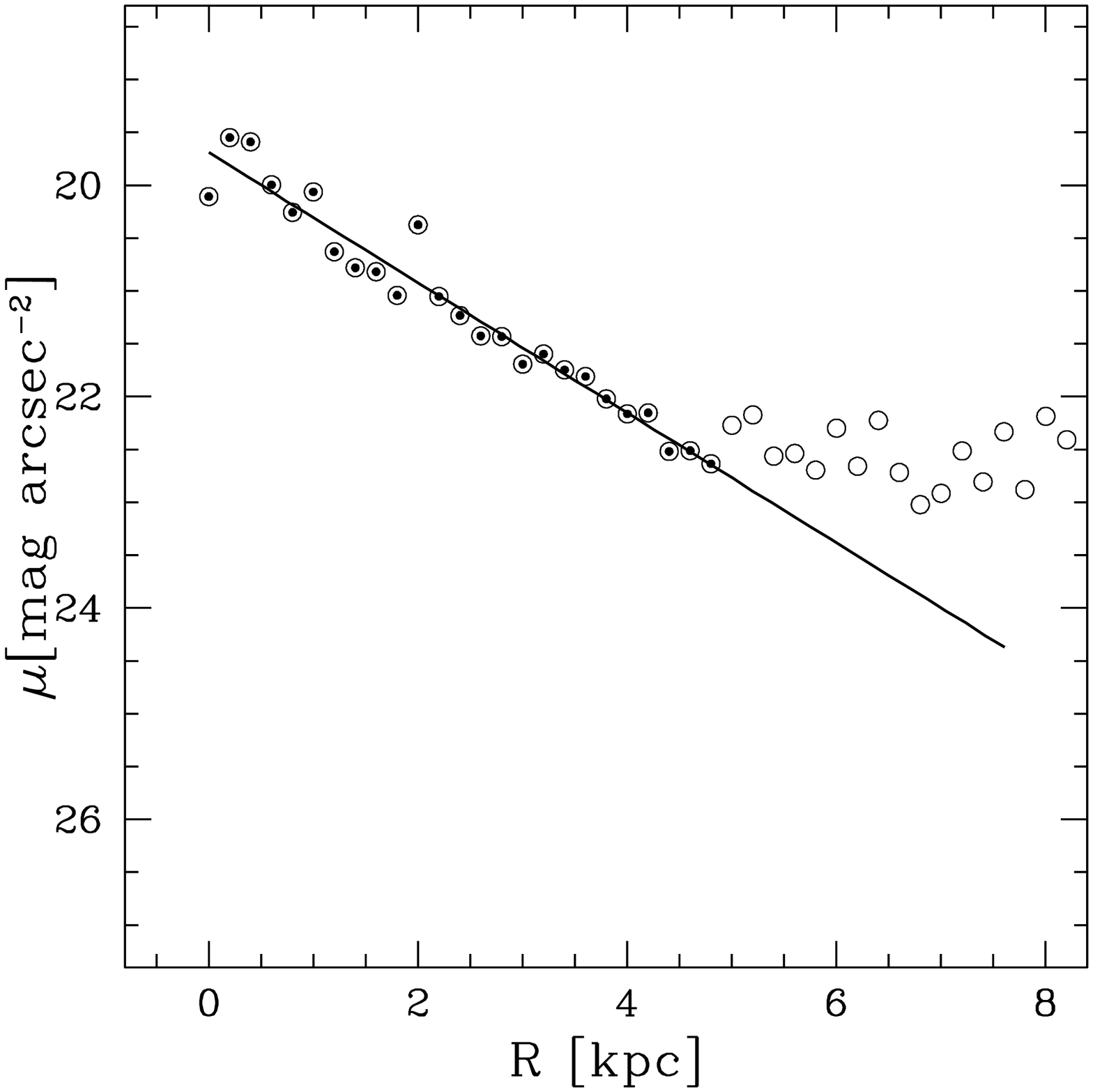}
\caption{Azimuthally averaged surface brightness profile (not corrected for inclination) 
of NGC 2366 in the 3.6 $\mu$m derived assuming the tilted-ring parameters shown in 
Fig. 9. The curve shown is fitted to the data at 
$R < 5$ kpc (partly filled circles).
\label{FIGURE16}}
\end{figure}
 {\clearpage}

\begin{figure}
\epsscale{1.0}
\plotone{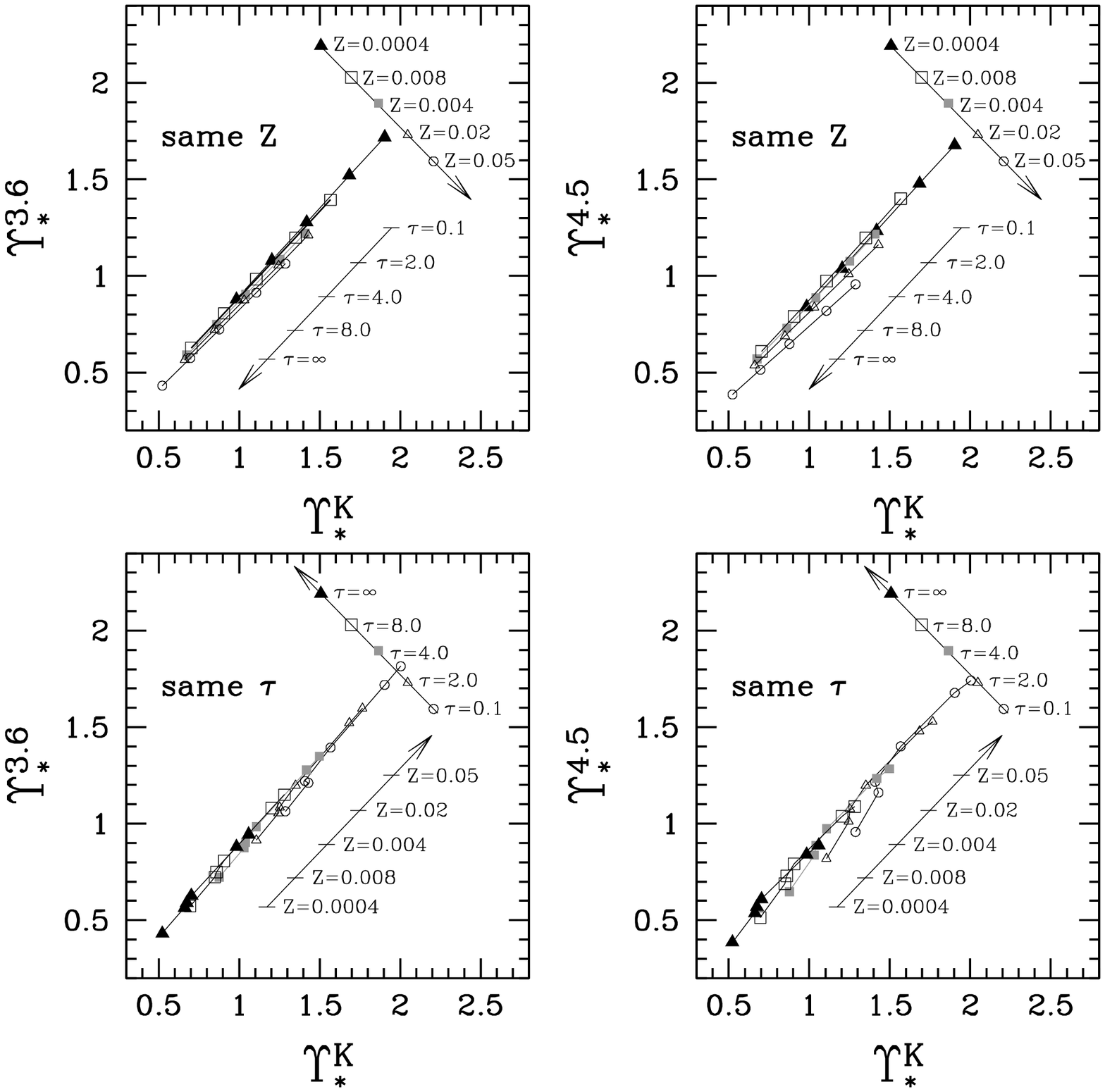}
\caption{The relations between the \MLk\ and the \ML\ in the 3.6 $\mu$m and 4.5 $\mu$m bands.
Metallicity is parameterized by the parameter $Z$ and $\tau$ represents the $e$-folding time-scale for the exponentially declining star formation (Bruzual 1983).
Upper panels: Models of the same metallicity $Z$ but different $\tau$ are connected by lines. 
$\tau$ ranges from 0.1 Gyr to $\infty$ as indicated by arrows.
$Z$ varies with 0.0004, 0.004, 0.008, 0.02, and 0.05 as illustrated by different symbols.
Lower panels: Models of the same $\tau$ but different metallicity $Z$ are connected by lines.
$Z$ ranges from 0.0004 to 0.05 as indicated by arrows. $\tau$ varies with 0.1, 2, 4, 8 Gyr,
and $\infty$ as illustrated by different symbols.
\label{FIGURE17}}
\end{figure}
 {\clearpage}

\begin{figure}
\epsscale{1.0}
\plotone{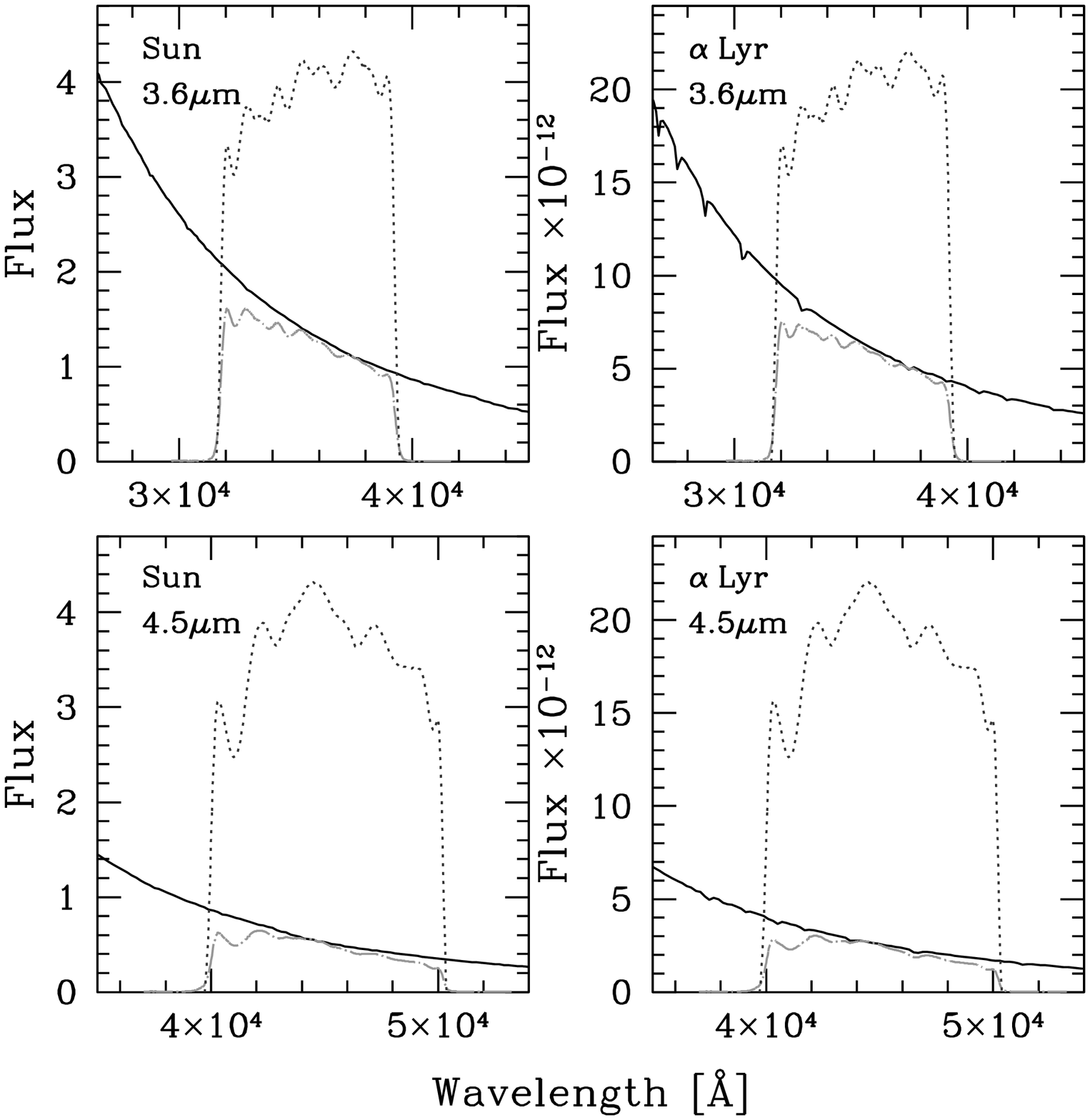}
\caption{The spectral energy distribution of the Sun and $\alpha$ Lyr convolved with the 
IRAC filter response functions for 3.6 $\mu$m and 4.5 $\mu$m. Flux is in units of 
erg cm$^{-2}$ s$^{-1}$ \AA$^{-1}$. Black solid-lines: Spectral energy distribution of the Sun and
$\alpha$ Lyr. Dotted lines: filter response functions of the IRAC bands (3.6 $\mu$m and 4.5 $\mu$m). Dash-dotted 
lines denote the convolved spectral energy distribution of the Sun and $\alpha$ Lyr.
\label{FIGURE18}}
\end{figure}
 {\clearpage}

\begin{figure}
\epsscale{1.0}
\plotone{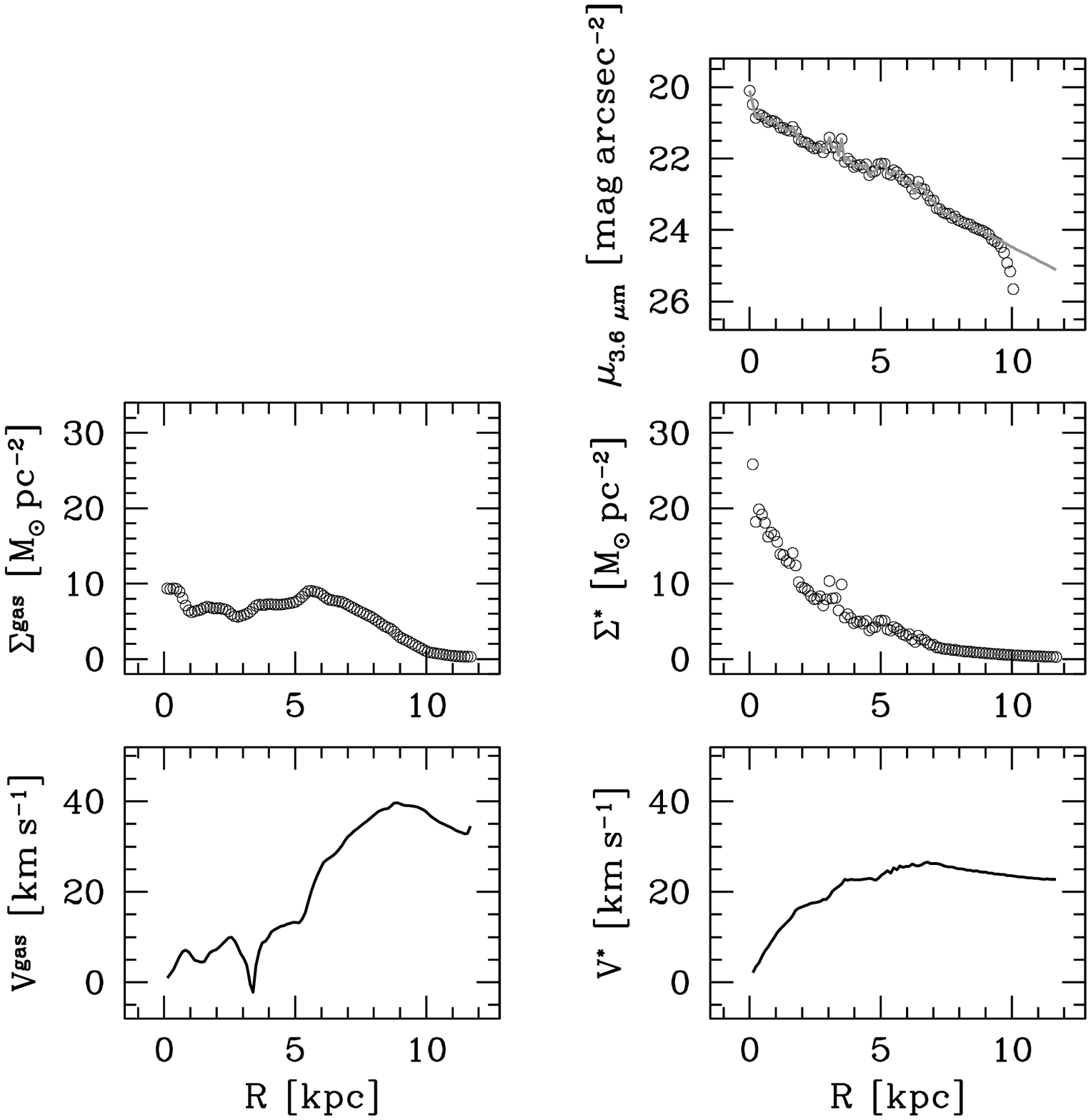}
\caption{Mass models for the gas and stellar components of IC 2574. The
  left two panels show the radial mass surface density distribution of
  neutral gas (observed H{\sc i} scaled by 1.4 to account for He and
  metals) and corresponding rotation curve derived from this.  The
  three panels on the right from top to bottom represent the radial
  surface brightness (inclination corrected) in the 3.6 $\mu$m band, stellar mass surface
  density, and rotation velocity of the disk using the model
  \MLsps\ values shown in Fig. 15. The radial average of
  the model \MLsps\ values is given in Table 3.
\label{FIGURE19}}
\end{figure}
 {\clearpage}

\begin{figure}
\epsscale{1.0}
\plotone{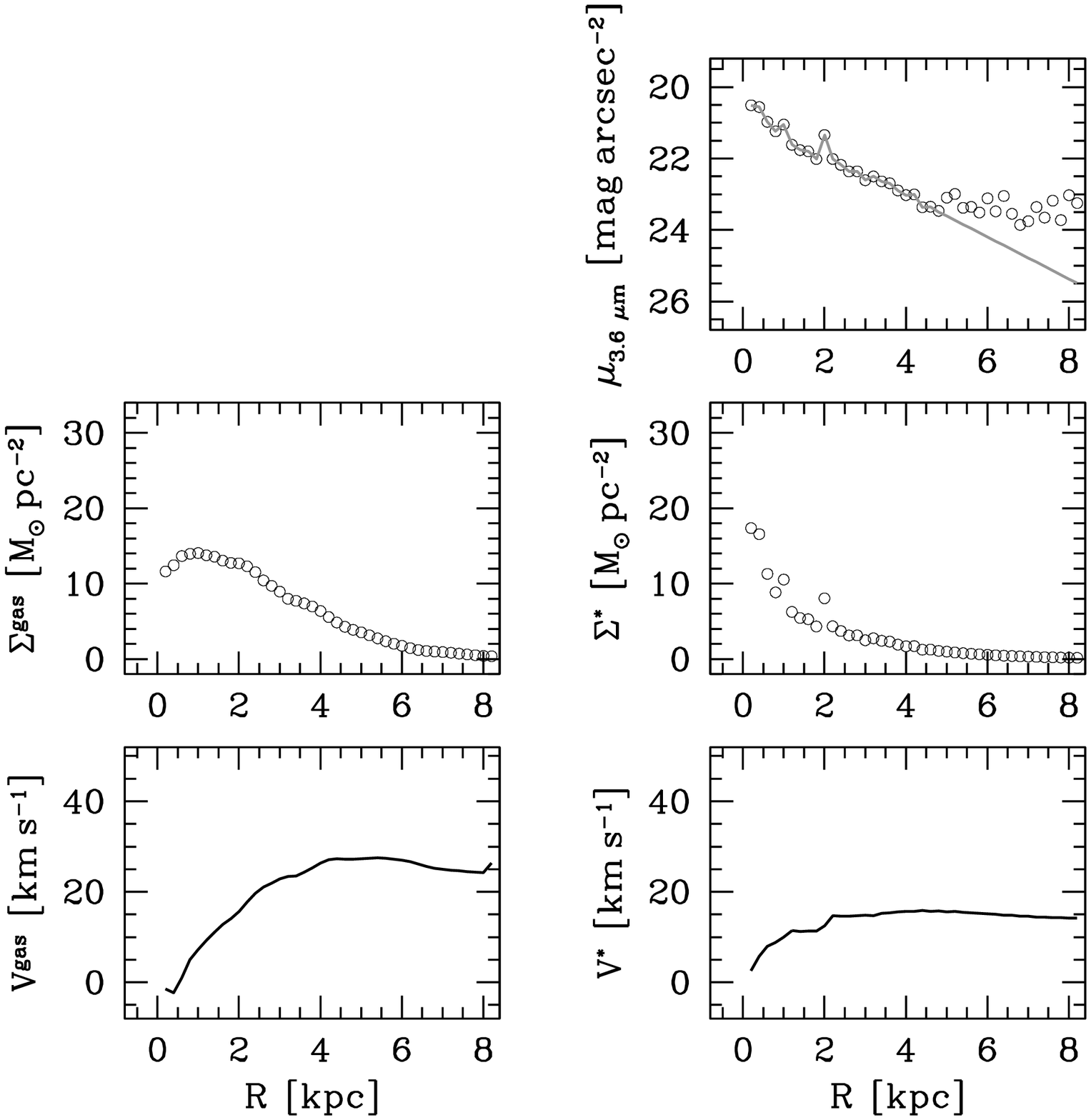}
\caption{Mass models for the gas and stellar components of
  NGC 2366. The left two panels show the radial mass surface density
  distribution of neutral gas (observed H{\sc i} scaled by 1.4 to
  account for He and metals) and corresponding rotation curve derived
  from this.  The three panels on the right from top to bottom
  represent the radial surface brightness (inclination corrected) in the 3.6 $\mu$m band,
  stellar mass surface density, and rotation velocity of the disk
  using the model \MLsps\ values given in Table 2.
\label{FIGURE20}}
\end{figure}
 {\clearpage}

\begin{figure}
\epsscale{1.02}
\plotone{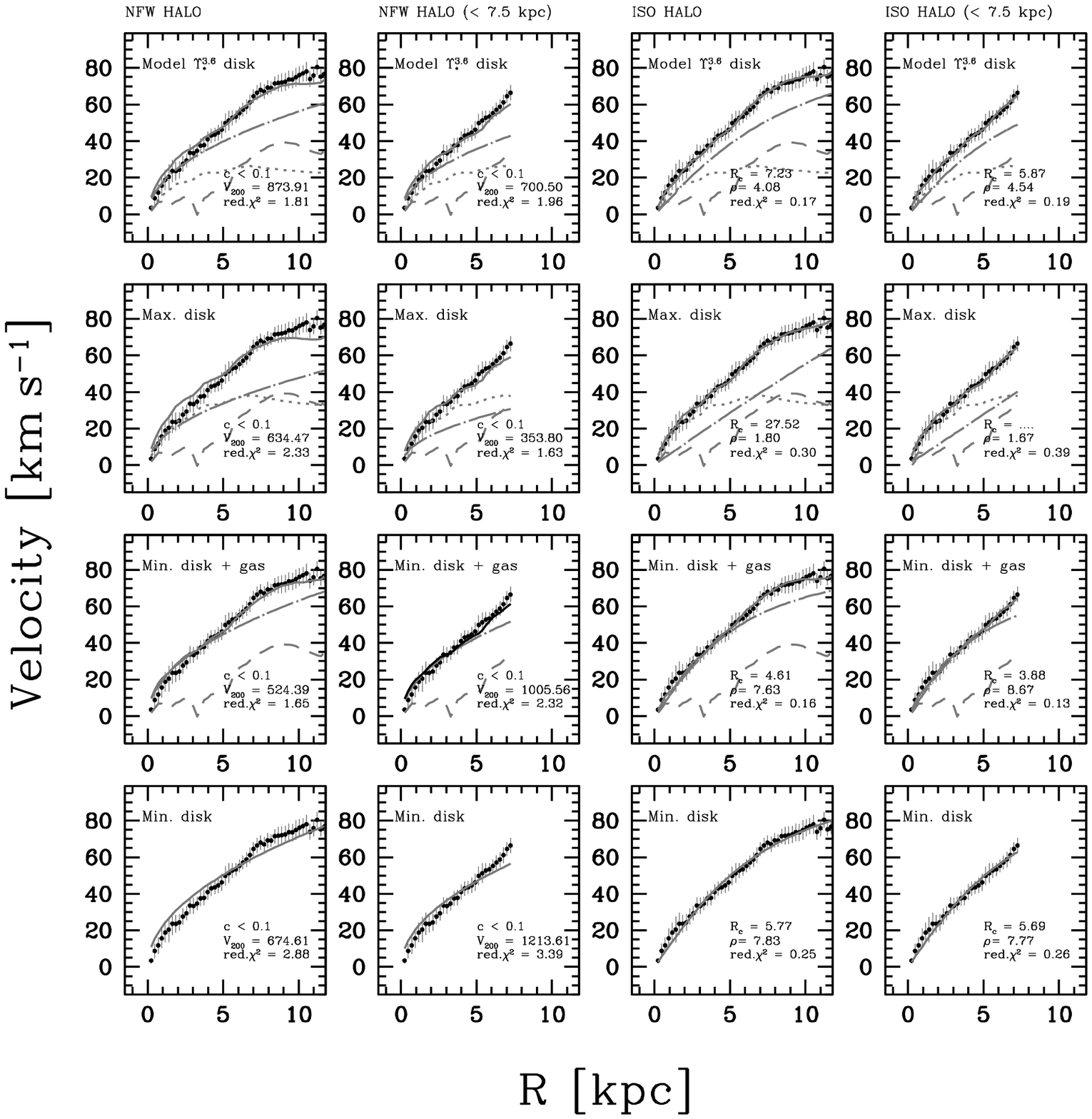}
\caption{Disk-halo decomposition of the IC 2574 rotation curve under various \ML\ assumptions 
(model \MLsps, maximum disk, minimum disk + gas, and minimum disk). 
The long dashed lines represent the rotation curves of the gas component; the short 
dashed lines are the rotation curves of the stellar disk; the long dash-dotted lines the 
rotation curves of the dark matter halo. The dots are the observed rotation curves from 
the bulk velocity field while the full lines are the sum of all contributions. 
The fitted parameters for each halo model (NFW and 
pseudo-isothermal dark matter halo models) are denoted on each panel along with the reduced $\chi^{2}$ 
value. Note that the pseudo-isothermal halo gives much better fits to the observed 
curves than the NFW halo. This is discussed more fully in Section 5.2.
\label{FIGURE21}}
\end{figure}
 {\clearpage}


\begin{figure}
\epsscale{1.02}
\plotone{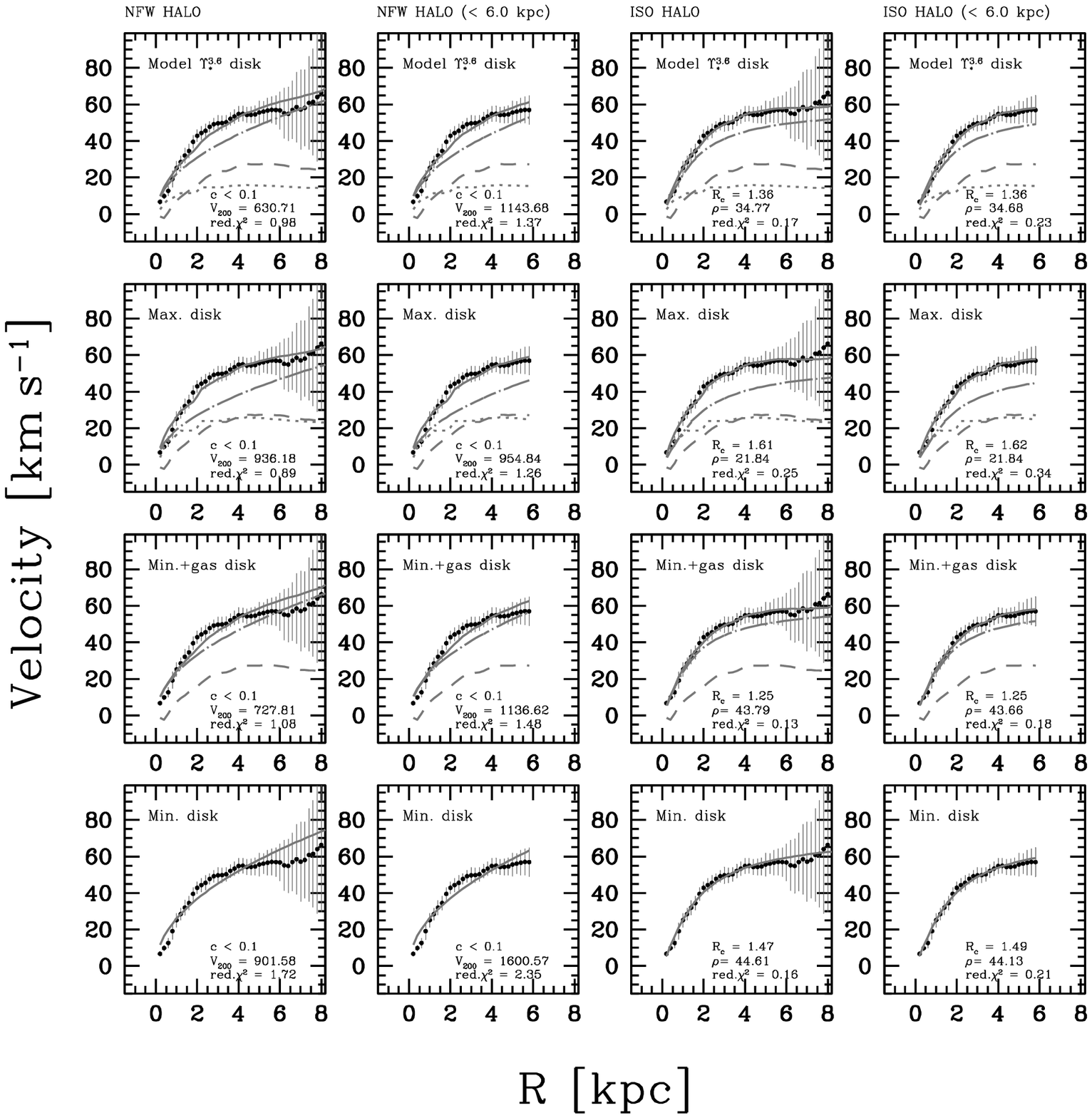}
\caption{Disk-halo decomposition of the NGC 2366 rotation curve under various \ML\ assumptions 
(model \MLsps, maximum disk, minimum disk + gas, and minimum disk). 
The long dashed lines represent the rotation curves of the gas component; the short 
dashed lines are the rotation curves of the stellar disk; the long dash-dotted lines the 
rotation curves of the dark matter halo. The dots are the observed rotation curves from 
the bulk velocity field while the full lines are the sum of all contributions.
The fitted parameters for each halo model (NFW and 
pseudo-isothermal dark matter halo models) are denoted on each panel along with the reduced $\chi^{2}$ 
value. Note that the pseudo-isothermal halo gives much better fits to the observed 
curves than the NFW halo. This is discussed in detail in Section 5.2.
\label{FIGURE22}}
\end{figure}
 {\clearpage}


\begin{figure}
\epsscale{1.0}
\plotone{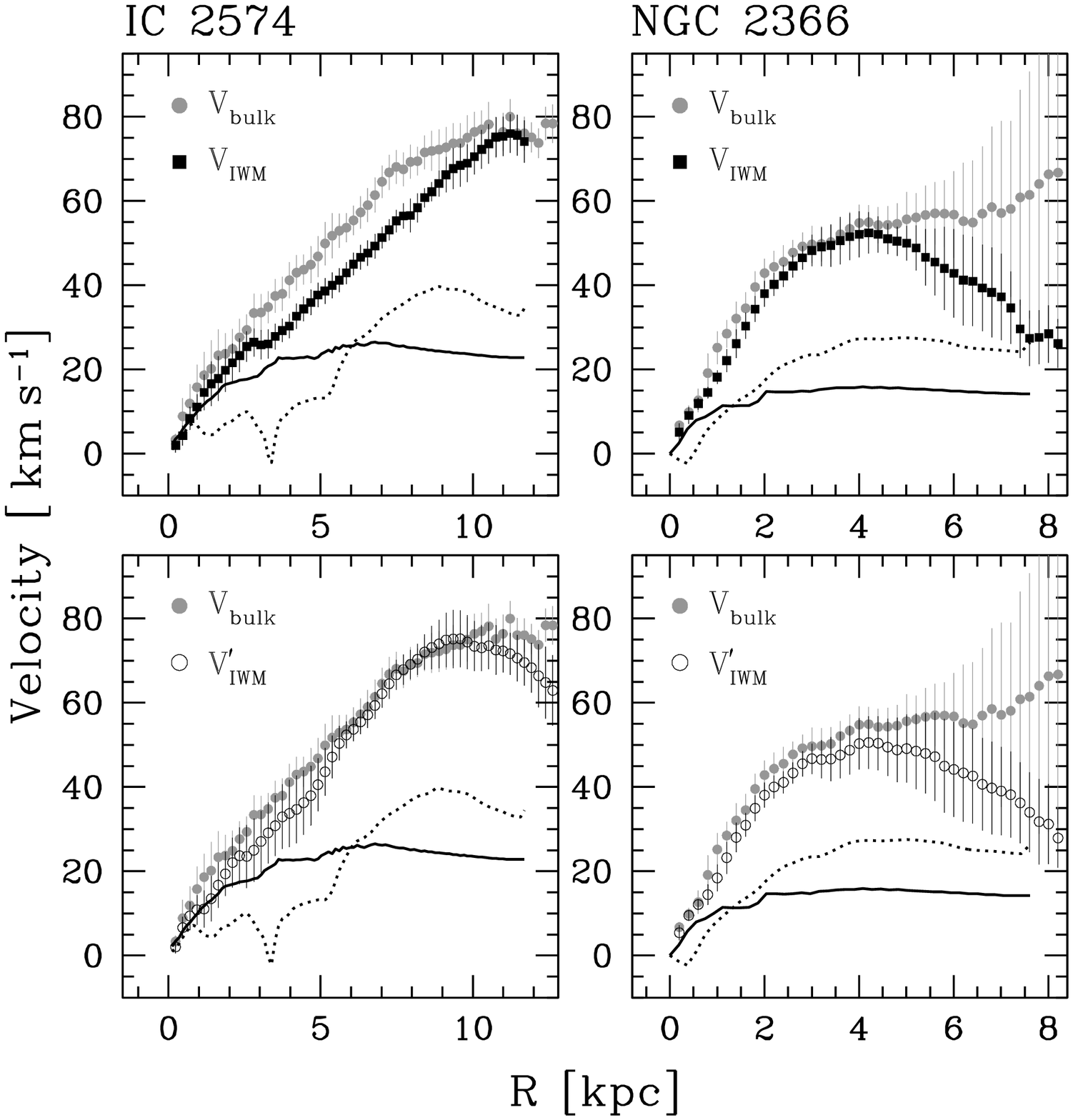}
\caption{Plots showing how non-circular motions affect the mass modeling for IC 2574 and NGC 2366.
Filled gray circles represent $V_{\rm bulk}$. Filled squares and open circles represent $V_{\rm IWM}$ and $V^{\prime}_{\rm IWM}$
respectively. The full lines are the contribution to the rotation curves of the stellar component; dotted lines represent
the contribution to the rotation curves of the gas component. 
The derived rotation velocity from the IWM velocity field, $V_{\rm IWM}$, which is disturbed by non-circular
motions, is underestimated. This results in the decreased contribution of 
dark matter to the observed kinematics since the stellar and gas components remain unchanged.
This is discussed in Section 5.2.
\label{FIGURE23}}
\end{figure}
 {\clearpage}


\begin{figure}[h!]
\epsscale{1.0}
\includegraphics[angle=0,width=0.95\textwidth,bb=5 420 565 780,clip=]{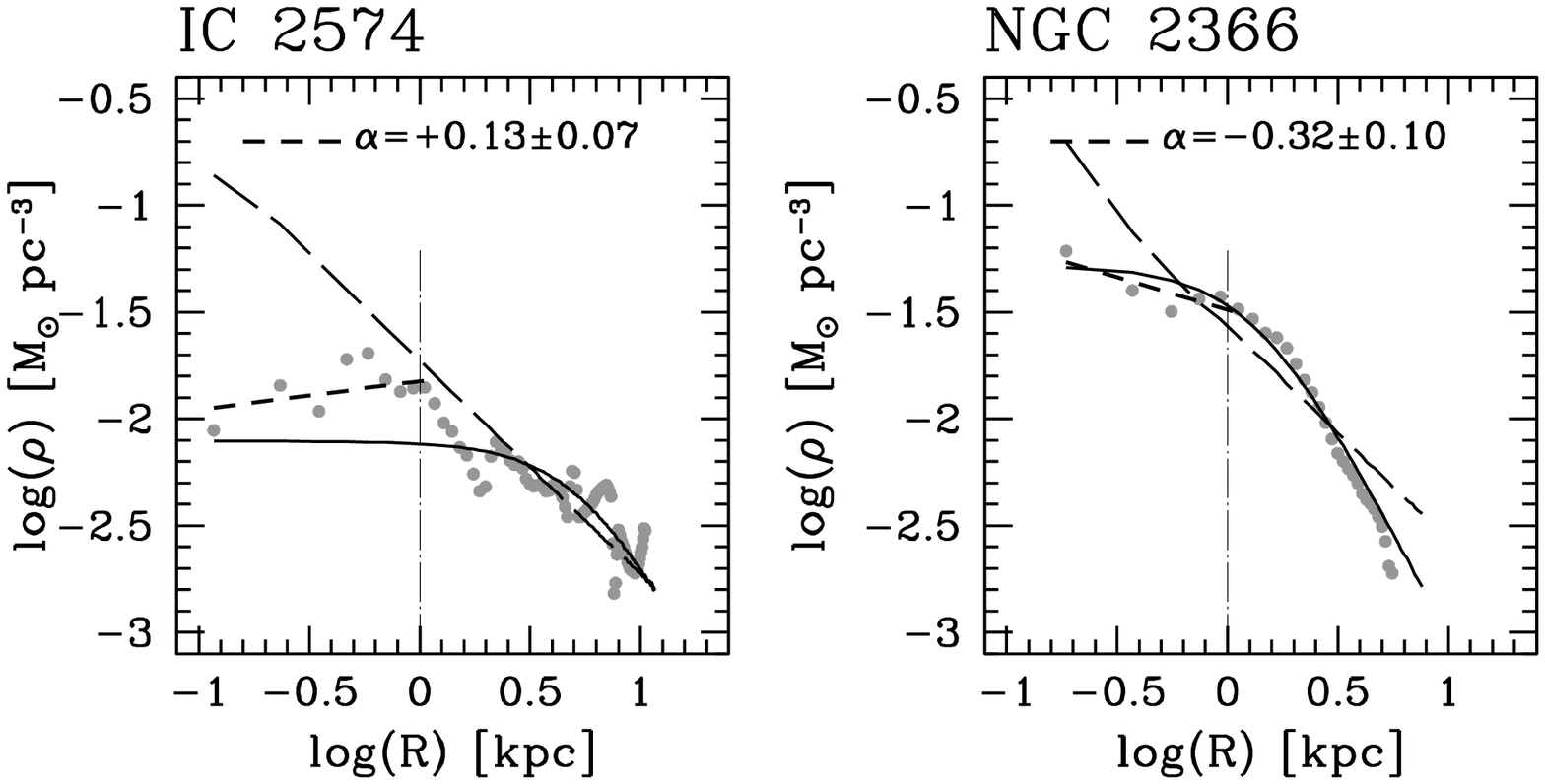}
\caption{The derived mass density profiles of IC 2574 and NGC 2366.
Long dashed and solid lines show the NFW halo model and
the pseudo-isothermal halo model, respectively.
Vertical long dash-dotted lines indicate 1 kpc radius.
The filled gray circles represent the dark matter
density profile derived from the bulk rotation velocity.
The inner slope of the derived dark matter density profile
is denoted by $\alpha$ and measured by a least squares fit 
(short dashed lines) to data points at radii 
less than 1.2 kpc.  The measured inner slopes of the mass density 
profiles of IC 2574 and NGC 2366 are shown in the panels.
\label{FIGURE24}}
\end{figure}
 {\clearpage}

\begin{figure}[h!]
\epsscale{1.0}
\plotone{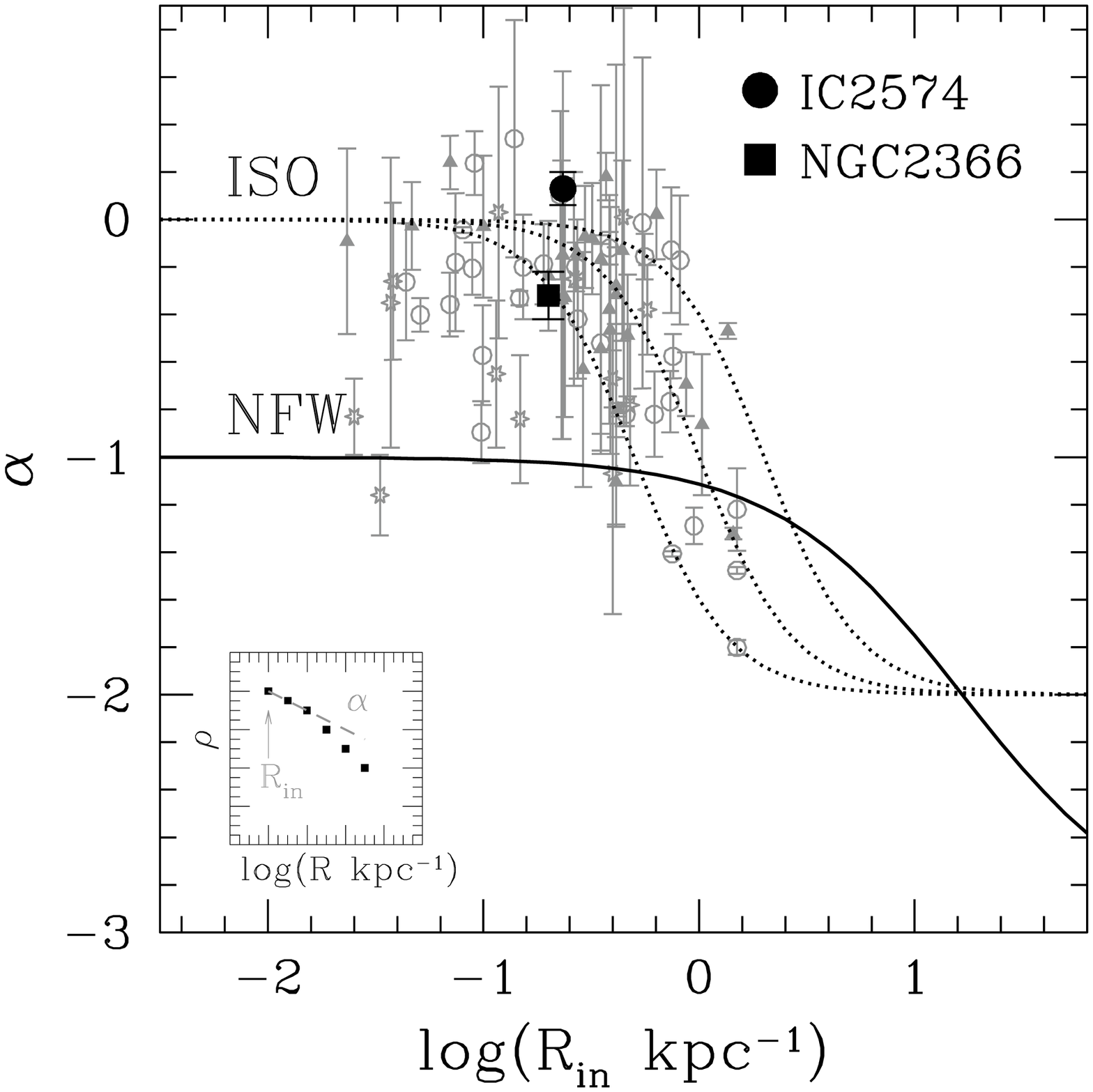}
\caption{The inner slope of the dark matter density profile plotted against the 
radius of the innermost point.
The inner-slopes of the mass density profiles of IC 2574 and NGC 2366 are overplotted with 
earlier work; they are consistent with previous measurements.
Open circles: de Blok \etal\ (2001); squares: de Blok \& Bosma (2002); open stars: Swaters \etal\ 
(2003).
The pseudo-isothermal model is preferred over the NFW model to explain the observational data.
\label{FIGURE25}}
\end{figure}
 {\clearpage}

\end{document}